%% file: main_arxiv.tex
\documentclass[%
 reprint,
superscriptaddress, 
 amsmath,amssymb,
 aps, pra
]{revtex4-2}

\usepackage{graphicx}
\usepackage{dcolumn}
\usepackage{bm}
\usepackage{dsfont}%
\usepackage{xcolor}
\usepackage{enumitem}
\usepackage{multirow}

\setlist[itemize]{align=parleft,left=0pt..5.5em}
\usepackage{hyperref}
\newcommand{\nn}{\nonumber}
\newcommand{\eqlab}[1]{\label{eq:#1}}
\renewcommand{\eqref}[1]{Eq.~(\ref{eq:#1})}
\newcommand{\eqsref}[2]{Eqs.~(\ref{eq:#1}) and~(\ref{eq:#2})}

\newcommand{\figref}[1]{Fig.~\ref{fig:#1}}

\newcommand{\figlab}[1]{\label{fig:#1}}
\newcommand{\tabref}[1]{Table~\ref{tab:#1}}

\newcommand{\tablab}[1]{\label{tab:#1}}
\newcommand{\appref}[1]{Appendix~\ref{app:#1}}

\newcommand{\applab}[1]{\label{app:#1}}
\newcommand{\secref}[1]{Section~\ref{sec:#1}}

\newcommand{\seclab}[1]{\label{sec:#1}}
\newcommand{\equal}{\!=\!}
\newcommand{\minus}{\!-\!}

\newcommand{\ket}[1]{|{#1}\rangle}
\newcommand{\bra}[1]{\langle {#1}|}
\newcommand{\expect}[1]{\big \langle {#1} \big \rangle}
\newcommand{\braket}[2]{\langle #1 | #2 \rangle}
\newcommand{\tauG}{\tau_{\!\scriptscriptstyle \mathcal{G}}}
\newcommand{\OmG}{\Omega_{\scriptscriptstyle \mathcal{G}}}

\newcommand{\Tend}{t_{\scriptscriptstyle N}}
\newcommand{\Tgate}{T}
\newcommand{\Fgate}{\mathcal{F}_{\rm{gate}}}
\newcommand{\chill}{\chi^{\scriptscriptstyle (2)} }
\newcommand{\chilll}{\chi^{\scriptscriptstyle (3)} }
\newcommand{\chiNL}{\Gamma_{\scriptscriptstyle \hspace{-0.8mm} \rm{NL}}}
\newcommand{\chiSHG}{\chi_{\scriptscriptstyle \rm{SHG}}}

\newcommand{\gammaDP}{\gamma_{\rm{dp}}}
\newcommand{\kappaL}{\kappa_{\hspace{-0.0mm} l}}
\newcommand{\QL}{Q_{\hspace{-0.2mm}\scriptscriptstyle L}}
\newcommand{\kappaC}{\kappa_{\hspace{-0.2mm}\scriptscriptstyle C}}

\newcommand{\zetaR}{\zeta_{\scriptscriptstyle R}}
\newcommand{\zetaI}{\zeta_{\scriptscriptstyle I}}

\begin{document}

\title{Controlled-Phase Gate by Dynamic Coupling of Photons to a Two-Level Emitter}

\author{Stefan Krastanov}
\affiliation{ Department of Electrical Engineering and Computer Science, Massachusetts Institute of Technology,
77 Massachusetts Avenue, Cambridge, Massachusetts 02139, USA}%

\author{Kurt Jacobs}
\affiliation{U.S. Army Research Laboratory, Sensors and Electron Devices Directorate, Adelphi, Maryland 20783, USA}%
\affiliation{Department of Physics, University of Massachusetts at Boston, Boston, Massachusetts 02125, USA}%

\author{Gerald Gilbert}
\affiliation{The MITRE Coorporation, 200 Forrestal Road, Princeton, New Jersey 08540, USA}%

\author{Dirk R. Englund}%
\affiliation{ Department of Electrical Engineering and Computer Science, Massachusetts Institute of Technology,
77 Massachusetts Avenue, Cambridge, Massachusetts 02139, USA}%

\author{Mikkel Heuck}
 \email{mheu@dtu.dk}
\affiliation{ Department of Electrical Engineering and Computer Science, Massachusetts Institute of Technology,
77 Massachusetts Avenue, Cambridge, Massachusetts 02139, USA}%
\affiliation{Department of Electrical and Photonics Engineering, Technical University of Denmark, 2800 Kgs.~Lyngby, Denmark}%

\date{\today}

\begin{abstract}
We propose an architecture for achieving high-fidelity deterministic quantum logic gates on dual-rail encoded photonic qubits by letting photons interact with a two-level emitter (TLE) inside an optical cavity. The photon wave packets that define the qubit are preserved after the interaction due to a quantum control process that actively loads and unloads the photons from the cavity and dynamically alters their effective coupling to the TLE. 
The controls rely on nonlinear wave mixing between cavity modes enhanced by strong externally modulated electromagnetic fields or on AC Stark shifts of the TLE transition energy.
We numerically investigate the effect of imperfections in terms of loss and dephasing of the TLE as well as control field miscalibration. Our results suggest that III-V quantum dots in GaAs membranes is a promising platform for photonic quantum information processing. 
\end{abstract}

\maketitle


\section{Introduction}

\input{./main_introduction}

\section{Equations of Motion \seclab{hej}}
\input{./main_equations_of_motion}
\section{Controlled-Phase Gate \seclab{controlled-phase gate}}
\input{./main_cphase_gate_application}

\section{Gate Performance \seclab{gate performance}}
\input{./main_gate_performance}

\section{Noise in the Control Fields}
\input{./main_noise_in_control_fields}

\section{Comparison of Nonlinearities}
\input{./main_nonlinearity_comparison}

\section{Discussion}
\input{./main_discussion}


\twocolumngrid 

\textit{Competing Interests}
The authors declare no competing interests. 

\textit{Data Availability}
All code used to solve and optimize the control master equations is available upon request, as well as the raw output of the simulation routines.

\textit{Author Contribution}
The control protocol was conceived by the authors through joint discussions. The detailed protocol simulation, optimization, and analysis was worked out by M. H. with help from S. K. The final manuscript was vetted by all authors.

\textit{Acknowledgments:} This work was partly funded by the AFOSR program FA9550-16-1-0391, supervised by Gernot Pomrenke (D. E.), the MITRE Quantum Moonshot Program (S. K., M. H., G. G., and D. E.), the ARL DIRA ECI grant \emph{``Photonic Circuits for Compact (Room-temperature) Nodes for Quantum Networks"} (K. J.), and the Villum Foundation program QNET-NODES grant no. 37417 (M.H.).

\bibliography{Mendeley}

\newpage
\clearpage
\onecolumngrid 
\appendix
\input{./app_rotating_frame}
\input{./app_fidelity_equation}
\input{./app_equations_of_motion}
\input{./app_dressed_state_picture}
\input{./app_dephasing}
\input{./app_FoM_comparison}

\end{document}

%% file: main_introduction.tex
In quantum networks, optical photons are the main carrier of quantum information.
The absence of direct interaction between photons and their high excitation energy make them immune to the otherwise pervasive thermal noise. 
Conversely, the lack of direct interaction creates significant challenges to the use of photons as the substrate for quantum computation, where fast, high-fidelity logic gates between the (photonic) qubits are necessary. 
Effective interactions derived from measurements~\cite{Knill2001} result in probabilistic gates. Instead, we focus on deterministic gate implementations through coherent photon-photon interactions based on optical nonlinearities. Bulk optical nonlinearities are attractive due to their potential for room temperature operation~\cite{Choi2017,Heuck2020,Heuck2020a,krastanov2021room,Li2020}, but their strength remains too weak. At cryogenic temperatures, stronger nonlinearities arise by coupling photons to ancillary quantum systems.
For instance, strong interactions between photons and two-level emitters (TLEs) have been realized in many physical systems including atoms~\cite{Rauschenbeutel1999, Birnbaum2005}, quantum dots~\cite{Englund2007, Hennessy2007}, molecules~\cite{Rattenbacher2019}, superconducting circuits~\cite{Wallraff2004}, and ions~\cite{Takahashi2020}. 
%
\begin{figure}[!t]
\centering
  \includegraphics[width=0.9\linewidth] {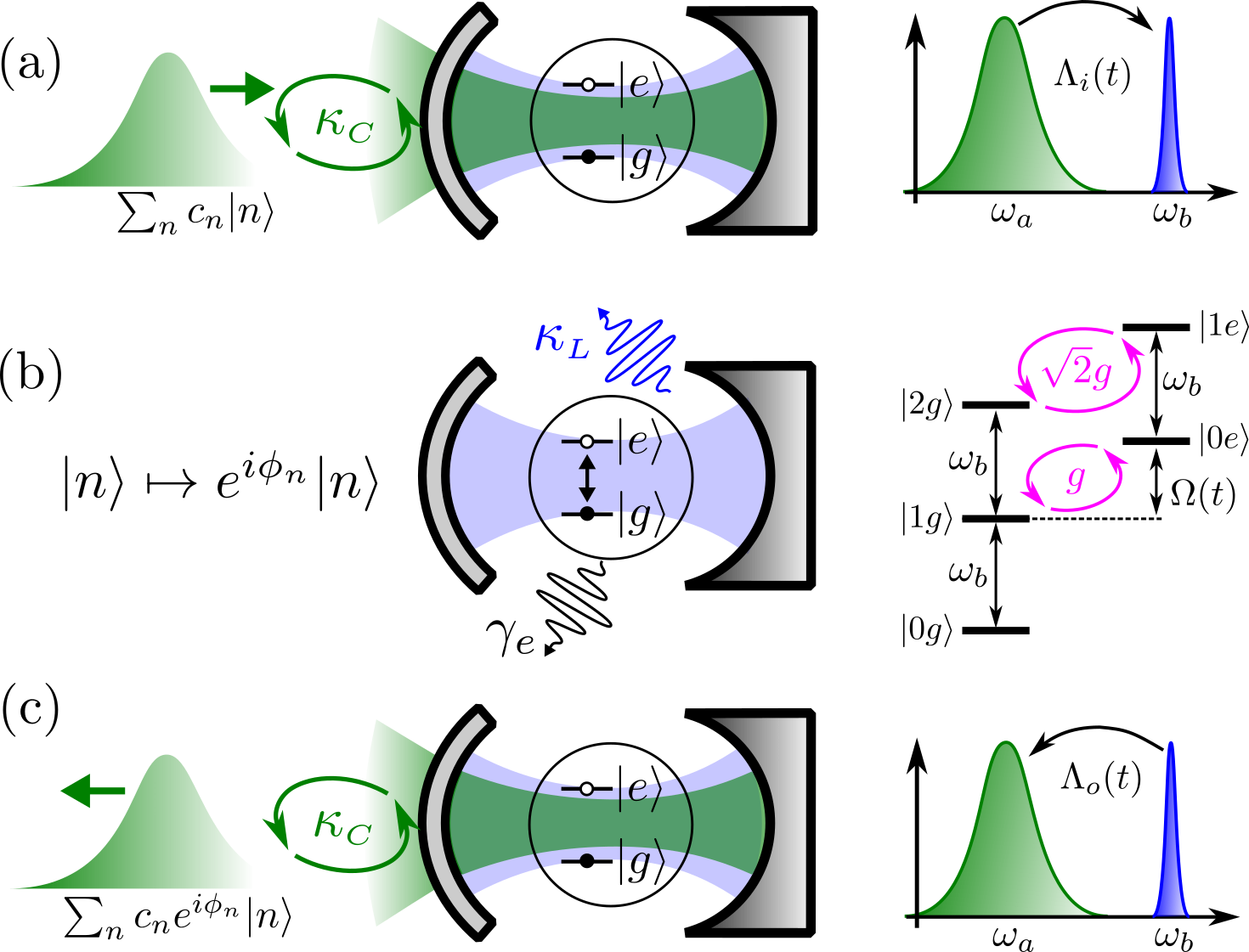} 
\caption{Schematic representation of the system and steps of the gate protocol.
(a) An incoming wave packet (green) in state $\sum c_n|n\rangle$ is coupled into a cavity mode $\hat{a}$ (also green). Mode $\hat{a}$ is coupled to a second available mode $\hat{b}$ (blue) via an external control field $\Lambda_i(t)$ as depicted in the spectrum diagram on the right. This coupling enables the capture of the incoming $\omega_a$ wave packet into mode $\hat{b}$.
(b) Mode $\hat{b}$ is coupled to a two-level emitter (TLE) at vacuum rate $g$. We control the detuning of the TLE through an external field described by $\Omega(t)$. As seen in the energy-level diagram, the cavity-TLE coupling depends on the number of photons in mode $\hat{b}$, which enables the depicted photon-number-dependent transformation.
(c) Controlled release using a second control pulse, $\Lambda_o(t)$. The state of the outgoing packet has undergone phase changes conditioned on the number of photons.}
\figlab{concept}
\end{figure} 
{{It is widely accepted that passive TLE-systems are insufficient to implement high-fidelity controlled-phase gates~\cite{Ralph2015, Nysteen2017}. Multi-stage approaches including active wave packet control~\cite{Ralph2015, Yang2022} increase resource overhead and optical loss. Ancillary qubits based on multi-level atomic systems~\cite{Duan2004,Johne2012,Iakoupov2018} provide added flexibility, but at a significant cost in technological complexity. A dynamic cavity control scheme was employed in Refs.~\cite{Heuck2020, Heuck2020a} for bulk nonlinearities, but it remains an open question whether a similar approach works for TLEs. The reason is that both one- and two-photon cavity states are straightforwardly coupled out once a $\pi$-phase difference between them is achieved~\cite{Heuck2020, Heuck2020a}. For TLEs, however, a state with $n$ photons has a Rabi frequency proportional to $\sqrt{n}$ so evacuating the cavity for both $n\equal 1$ and $n\equal 2$ is nontrivial.

Here, we introduce a single-stage dynamic control scheme for photonic qubits that exploits the strong interactions with a TLE in a multimode cavity.~\figref{concept} illustrates how photons travelling in wave packets are actively loaded into a resonator where they interact via the TLE and are subsequently released into the same wave packet with transformed photon-number contents. We assume to have control over the detuning between the TLE and cavity mode $\hat{b}$ such that $\Omega(t)\equal \omega_e-\omega_b$. This provides control over the effective Rabi frequency $\sqrt{g^2+\Omega(t)^2/4}$ (see~\appref{dressed state picture}) to enable both one- and two-photon input states to be coupled out efficiently. We also assume to have control over the coupling between cavity modes $\hat{a}$ and $\hat{b}$ with a rate $\Lambda(t)$. This can be achieved by three-wave-mixing between the aforementioned two modes and a strong controlled classical pump~\cite{McCutcheon2009, Heuck2020}. Note that Ref.~\cite{Choi2019} similarly used two cavity modes coupled by a time-\emph{independent} rate to improve the trade-off between indistinguishability and efficiency of a quantum emitter.

Since the time-dependent cavity-TLE detuning effectively controls the strength of the nonlinearity, the gate duration can be shortened without reducing the fidelity, in contrast to passive nonlinearities~\cite{Heuck2020, Heuck2020a}. By numerical optimization of $\Omega(t)$ and $\Lambda(t)$, we show that high-fidelity controlled-phase gates are, in fact, possible and further that the gate duration need only exceed the Rabi period at zero-detuning by a factor of 2-3. }}

This manuscript is organized as follows: In the next section we derive the general form of the equations of motion for one- and two-photon wave packets incident on the cavity-TLE system. This serves at the basis for our control conditioned on the photon number of the wave packets. These equations are used in Section III to derive control fields that enable a high-fidelity controlled-phase gate as an example of the many logical operations enabled by this design. Section IV provides a detailed study of the performance of that gate with respect to various hardware parameters. Lastly, we provide an outlook on possible near-term hardware implementations and concluding remarks.

%% file: main_equations_of_motion.tex
Before demonstrating the implementation of a controlled-phase gate, we will describe the general form of the dynamics of capturing (and releasing) a wave packet into our two-mode cavities in the presence of a TLE. The TLE is crucial for the non-Gaussian quantum operations we want to perform.
We use the discrete-time formalism developed in Ref.~\cite{Heuck2020} to describe the system in~\figref{concept}. It involves discretizing the time axis into $N$ bins of width $\Delta t$ and introducing discrete-time waveguide mode operators 
\begin{align}\eqlab{discrete field operator}
   \hat{w}(t_k) = \hat{w}(k\Delta t) \equiv \frac{\hat{w}_k}{\sqrt{\Delta t}}  \;\;\; \mbox{with} \;\;\;  [\hat{w}_j,\hat{w}_k^\dagger] = \delta_{jk} ,  
\end{align}
where $\hat{w}(t_k)$ is the continuous-time annihilation operator of the waveguide mode that couples to cavity mode $\hat{a}$. The input state of a single photon is 
\begin{align}\eqlab{discrete single photon input state}
    \ket{\psi_{\rm{in}}^{\scriptscriptstyle (1)}} = \int_{t_0}^{\Tend} \!\!dt \xi_{\rm{in}}(t) \hat{w}^\dagger(t) \ket{\emptyset} \approx \sum_{k=1}^N \sqrt{\Delta t} \xi^{\rm{in}}_k  \hat{w}_k^\dagger \ket{\emptyset}, 
\end{align}
where $\int_{t_0}^{\Tend}|\xi_{\rm{in}}(t)|^2 = 1$ so the state is normalized, $\xi^{\rm{in}}_k\equal \xi_{\rm{in}}(t_k)$ describes the shape of the wave packet, and $\ket{\emptyset}$ denotes the vacuum state of the waveguide. To each time bin, $n$, there is an associated Hamiltonian 
\begin{multline}\eqlab{interaction picture Hamiltonian}
    \frac{\hat{H}_{n}}{\hbar} =   \Omega_n\hat{\sigma}_{ee} + i \sqrt{\frac{\kappaC}{\Delta t}} \Big( \hat{a}^\dagger \hat{w}_n - \hat{a}\hat{w}_n^\dagger \Big) ~+\\
    \Lambda^{\!*}_n\hat{a}^\dagger\hat{b} +\Lambda_n\hat{a}\hat{b}^\dagger  + g\Big(\hat{b}^\dagger \hat{\sigma}_{-}  + \hat{b} \hat{\sigma}_{+} \Big),
\end{multline} 
which describes the interaction between the cavity and photons in the waveguide at time $t_n$ as well as the internal dynamics of the cavity. The propagation of the wave packet is, thus, handled implicitly (instead of introducing an additional hopping Hamiltonian). The operators describing the TLE in~\eqref{interaction picture Hamiltonian} are $\hat{\sigma}_{ee} = \ket{e}\bra{e}$, $\hat{\sigma}_{-} = \ket{g}\bra{e}$, and $\hat{\sigma}_{+} = \ket{e}\bra{g}$. {The coupling rate between cavity mode $\hat{a}$ and the waveguide is $\kappaC$, the controllable coupling between modes $a$ and $b$ is $\Lambda_n$, and the coupling rate between the emitter and photons in mode $\hat{b}$ is $g$}. Note that the Hamiltonian in~\eqref{interaction picture Hamiltonian} corresponds to a rotating frame as described in~\appref{rotating frame}. Photons in any time bin only interact with the cavity once and the bins are ordered such that photons in the first bin interact with the cavity first. At a given time step, $t_n$, we therefore denote all photons in bins $t_k>t_n$ as \textit{input} photons and write their state as $\ket{1_k}$. Similarly, photons in all bins after the cavity-interaction $t_k\leq t_n$ are denoted \textit{output} photons and their state is written in bold as $\ket{\bm{1}_k}$. The state of the cavity-TLE system is $\ket{n_a n_b g}$ or $\ket{n_a n_b e}$ when there are $n_a$ photons in mode $\hat{a}$ and $n_b$ photons in mode $\hat{b}$ while the TLE is in the ground, $\ket{g}$, or excited state, $\ket{e}$, respectively.

The flow diagram in~\figref{input-output map} maps out the various paths that two input photons may take while interacting with the system. Each arrow corresponds to a non-zero coupling in the system. We use the diagram as a visual tool to simplify the otherwise tedious job of writing down the dynamical equations.
\begin{figure}
\centering
  \includegraphics[width=8.0cm] {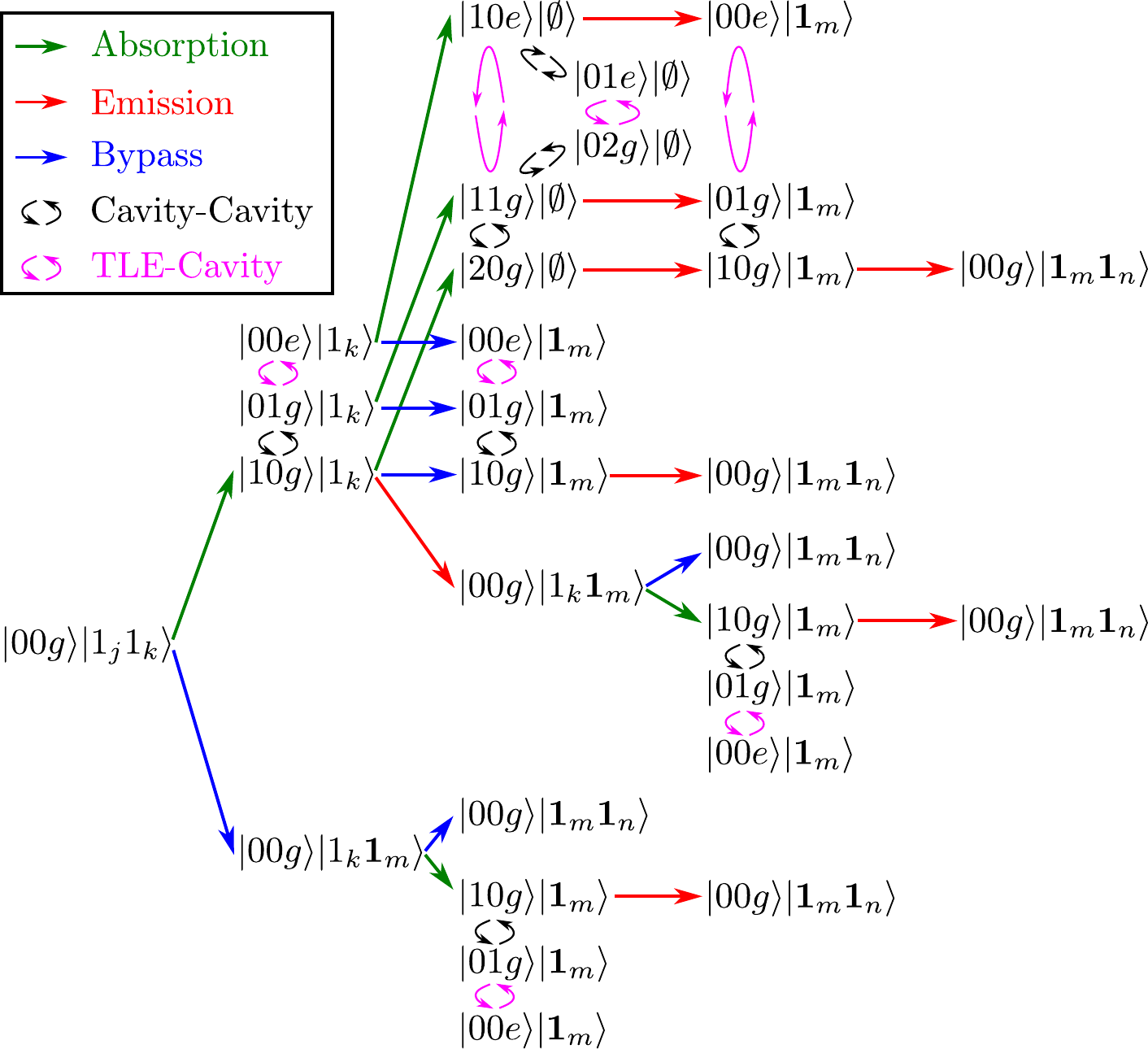} 
\caption{All input-output paths for two incident photons. Each photon may be absorbed (green) and re-emitted (red) or bypass the cavity by reflecting off of the left mirror in~\figref{concept}(a) at time $t_m$. Inside the cavity, the photons may couple between the cavity modes (black) or between the TLE and mode $\hat{b}$ (magenta). }
\figlab{input-output map}
\end{figure} 
The Schr\"odinger coefficients corresponding to states with up to two photons remaining on the input side of the cavity at time $t$ are denoted e.g. $\psi_{n_a n_b g}^{\scriptscriptstyle (2)} (t)$, where the superscript denotes the maximum number of input photons. As an example, consider the state if the first photon is absorbed into mode $\hat{a}$ and subsequently coupled to mode $\hat{b}$ before the second photon reaches the cavity. The corresponding state is $\psi_{01g}^{\scriptscriptstyle (2)} (t) \ket{01g}\ket{1_k}$ with $t_k>t$. States with a maximum of one photon remaining on the input side along with an output photon in bin $m$ have Schr\"odinger coefficients denoted e.g. $\psi_{n_a n_b g}^{\scriptscriptstyle (1)} (t_m, t)$. States with no photons on the input side and an output photon in bin $m$ have coefficients $\psi_{n_a n_b g}^{\scriptscriptstyle (0)} (t_m, t)$. Finally, coefficients corresponding to states with both photons in the cavity-TLE system are denoted e.g. $\psi_{n_a n_b g}(t)$ without a superscript.

We refer to Ref.~\cite{Heuck2020} for details of deriving equations of motion for the Schr\"odinger coefficients, input-output relations, and inclusion of loss channels. For the coefficients mentioned above, describing the state of the cavity in the aforementioned basis, the master equation results in
\begin{subequations}\eqlab{eoms two modes two photons}
\begin{align}
    \dot{\psi}_{10g}^{\scriptscriptstyle (L)}( t_m ,t)  =  &-\frac{ \kappa}{2}\psi_{10g}^{\scriptscriptstyle (L)}( t_m ,t) - i\Lambda(t)^{\!*}\psi_{01g}^{\scriptscriptstyle (L)}( t_m ,t) ~\nn\\
    &+\sqrt{L}\sqrt{\kappaC}\xi_{\rm{in}}(t) \eqlab{eoms psi10gM}\\
    \dot{\psi}_{01g}^{\scriptscriptstyle (L)}( t_m ,t)  = &-\frac{ \kappaL}{2} \psi_{01g}^{\scriptscriptstyle (L)}( t_m ,t) - i\Lambda(t)\psi_{10g}^{\scriptscriptstyle (L)}( t_m ,t) ~\nn\\
    &- i g \psi_{00e}^{\scriptscriptstyle (L)}( t_m ,t) \eqlab{eoms psi01gM}\\
    \dot{\psi}_{00e}^{\scriptscriptstyle (L)}( t_m ,t)  = &-\Big(i\Omega(t)  + \frac{\gamma_e}{2} \Big)\psi_{00e}^{\scriptscriptstyle (L)}( t_m ,t) ~\nn\\
    &- i g\psi_{01g}^{\scriptscriptstyle (L)}( t_m ,t) \eqlab{eoms psi00eM}.
\end{align} 
\end{subequations}
The first equation describes the capture of an incoming photon in cavity mode $\hat{a}$ and the interaction between $\hat{a}$ and $\hat{b}$. The latter equations introduce the interaction between mode $\hat{b}$ and the TLE. Note that we included a loss rate, $\kappaL$, for both cavity modes and a decay rate, $\gamma_e$, from the TLE to the electromagnetic environment. The total intensity decay rate from cavity mode $\hat{a}$ is $\kappa = \kappaC + \kappaL$ in~\eqref{eoms psi10gM} and $L=(0,1,2)$ is the maximum number of photons on the input side as described above. Note that solving~\eqref{eoms two modes two photons} with $L\equal 1$ and $t_m\equal 0$ corresponds to a one-photon input state. 

The Schr\"odinger coefficients corresponding to both photons being in the TLE-cavity system evolve according to 
\begin{subequations}\eqlab{ODEs two photon terms} 
\begin{align}
    \!\dot{\psi}_{20g}(t)  = &-\kappa \psi_{20g}(t) - i\sqrt{2}\Lambda(t)^{\!*}\psi_{11g}(t) ~\nn\\
    &+ \sqrt{2\kappaC}\psi_{10g}^{\scriptscriptstyle (2)}(t)\xi_{\rm{in}}(t) \\
    \!\dot{\psi}_{11g}(t)  = &- \frac{\kappaC+2\kappaL}{2} \psi_{11g}(t) ~\nn\\
    & - i\sqrt{2}\Big[\Lambda(t)\psi_{20g}(t) + \Lambda(t)^{\!*}\psi_{02g}(t)\Big]  ~\nn\\
    &-i g \psi_{10e}(t) + \sqrt{\kappaC}\psi_{01g}^{\scriptscriptstyle (2)}(t)\xi_{\rm{in}}(t) \\
    \!\dot{\psi}_{02g}(t)  = &- \kappaL\psi_{02g}(t) - i\sqrt{2}\Lambda(t)\psi_{11g}(t)  \nn\\
    &-i\sqrt{2} g \psi_{01e}(t) \eqlab{eoms psi02g}\\
    \!\dot{\psi}_{10e}(t)  = &-\!\Big(i\Omega(t)+\frac{\gamma_e+\kappa}{2} \Big)\psi_{10e}(t) - i\Lambda(t)^{\!*}\psi_{01e}(t) \nn\\
    &-ig\psi_{11g}(t) + \sqrt{\kappaC}\psi_{00e}^{\scriptscriptstyle (2)}(t)\xi_{\rm{in}}(t) \\
    \!\dot{\psi}_{01e}(t)  = &-\Big(i\Omega(t)+\frac{\gamma_e}{2}+\frac{\kappaL}{2} \Big)\psi_{01e}(t) - i\Lambda(t)\psi_{10e}(t) \nn\\
    &-i\sqrt{2} g \psi_{02g}(t) \eqlab{eoms psi01e}. 
\end{align}  
\end{subequations}
The initial condition for~\eqref{ODEs two photon terms} is that all coefficients are zero at $t=t_0$. Note that all driving terms in~\eqsref{eoms two modes two photons}{ODEs two photon terms} correspond to green arrows in~\figref{input-output map} while all terms proportional to $\Lambda$ and $g$ correspond to black and magenta arrows, respectively. As such,~\figref{input-output map} provides a convenient tool for verifying that all interactions are included in the dynamical equations.

For $L=2$ in~\eqref{eoms two modes two photons}, the only required initial condition is: $\psi_{10g}^{\scriptscriptstyle (2)} (0) = \psi_{01g}^{\scriptscriptstyle (2)} (0) = \psi_{00e}^{\scriptscriptstyle (2)} (0) = 0$. Those coefficients are therefore only functions of a single variable, $t$. For $L=1$ and $L=0$, the equations must be solved for $N$ different initial conditions since $t_m$ corresponds to any bin and the coefficients are functions of both $t_m$ and $t\geq t_m$. 

For $L=1$, the dynamics is initiated by either an emission into the waveguide or simply a bypass (the traveling photon passing by the cavity)
\begin{subequations}\eqlab{psi^(1) ini}
\begin{align}
    \ket{10g}\ket{1_k} &~{\color{red} \rightarrow}~\ket{00g}\ket{1_k \bm{1}_m}  \\
    \ket{00g}\ket{1_j 1_k} &~{\color{blue} \rightarrow}~\ket{00g}\ket{1_k \bm{1}_m}.
\end{align} 
\end{subequations}
In either case, the initial conditions are: $\psi_{10g}^{\scriptscriptstyle (1)} (t_m, t_m) = \psi_{01g}^{\scriptscriptstyle (1)} (t_m, t_m) = \psi_{00e}^{\scriptscriptstyle (1)} (t_m, t_m) = 0$. 

For $L=0$, the dynamics is initiated by one of three different emission paths or three different bypass paths
\begin{subequations}\eqlab{psi^(0) ini}
\begin{align}
    \ket{10e}\ket{\emptyset} &~{\color{red} \rightarrow}~\ket{00e}\ket{\bm{1}_m}\eqlab{psi^(0) ini a}\\
    \ket{11g}\ket{\emptyset} &~{\color{red} \rightarrow}~\ket{01g}\ket{\bm{1}_m}\\
    \ket{20g}\ket{\emptyset} &~{\color{red} \rightarrow}~\ket{10g}\ket{\bm{1}_m}\\
    \ket{00e}\ket{1_k} &~{\color{blue} \rightarrow}~\ket{00e}\ket{\bm{1}_m}\\
    \ket{01g}\ket{1_k} &~{\color{blue} \rightarrow}~\ket{01g}\ket{\bm{1}_m}\\
    \ket{10g}\ket{1_k} &~{\color{blue} \rightarrow}~\ket{10g}\ket{\bm{1}_m}.
\end{align} 
\end{subequations}
To understand how to set the initial conditions of~\eqref{eoms two modes two photons} with $L\equal 0$ based on the events listed in~\eqref{psi^(0) ini}, we consider the entire paths through the map in~\figref{input-output map}. As an example, we consider the top path, which~\eqref{psi^(0) ini a} is a part of
\begin{multline}\eqlab{path example}
    \ket{00g}\ket{1_j 1_k} ~{\color{green} \rightarrow}~ \ket{10g}\ket{1_k} ~{\color{green} \rightarrow}~ \ket{10e}\ket{\emptyset} ~{\color{red} \rightarrow}~ \\
    -\sqrt{\kappaC\Delta t}\psi_{10e}(t_m) \ket{00e}\ket{\bm{1}_m} ~{\color{red} \rightarrow}~\\
    -\sqrt{\kappaC\Delta t}\psi_{10g}^{\scriptscriptstyle (0)}(t_m, t_n) \ket{00g}\ket{\bm{1}_m \bm{1}_n} .
\end{multline} 
At each emission or bypass event, we explicitly write out the coefficient of the relevant state and the initial condition of~\eqref{eoms two modes two photons} is therefore $\psi_{00e}^{\scriptscriptstyle (0)}(t_m,t_m)\equal -\sqrt{\kappaC\Delta t}\psi_{10e}(t_m)$, while the other coefficients are initialized with the value zero. However, since~\eqref{eoms two modes two photons} is linear, we may use the initial value 1 and multiply the contributions to the output state in the end, such that the contribution from the path in~\eqref{path example} is $\kappaC\psi_{10e}(t_m)\psi_{10g}^{\scriptscriptstyle (0)}(t_m, t_n)$. To distinguish between which of the three coefficients $\psi_{00e}^{\scriptscriptstyle (0)}$, $\psi_{01g}^{\scriptscriptstyle (0)}$, or $\psi_{10g}^{\scriptscriptstyle (0)}$ that is initialized to 1, we define functions $\{A_{00e},A_{01g}, A_{10g}\}$, $\{B_{00e},B_{01g}, B_{10g}\}$, and $\{C_{00e},C_{01g}, C_{10g}\}$, where $A$ corresponds to $\psi_{10g}^{\scriptscriptstyle (0)}(t_m,t_m)\equal 1$, $B$ to $\psi_{01g}^{\scriptscriptstyle (0)}(t_m,t_m)\equal 1$, and $C$ to $\psi_{00e}^{\scriptscriptstyle (0)}(t_m,t_m)\equal 1$. Following all the paths in~\figref{input-output map}, the output state is found to consist of the following ten terms
\begin{subequations}\eqlab{input output two photons} 
\begin{multline}
    \xi_{\rm{out}}^{\scriptscriptstyle (2)}(t_m, t_n) = \frac{\kappaC}{\sqrt{2}}\bigg[\psi_{10e}(t_m) C_{10}(t_m, t_n)~\\
    +\psi_{11g}(t_m) B_{10}(t_m, t_n) + \sqrt{2} \psi_{20g}(t_m) A_{10}(t_m, t_n) ~\\
    - \frac{\xi_{\rm{in}}(t_m)}{\sqrt{ \kappaC}} \bigg(\! \psi_{00e}^{\scriptscriptstyle (2)}(t_m)C_{10}(t_m, t_n) ~\\
    + \psi_{01g}^{\scriptscriptstyle (2)}(t_m)B_{10}(t_m, t_n) +\psi_{10}^{\scriptscriptstyle (2)}(t_m)A_{10}(t_m, t_n) \!\bigg) ~\\
    - \frac{\psi_{10g}^{\scriptscriptstyle (2)}(t_m)}{\sqrt{ \kappaC}} \bigg( \xi_{\rm{in}}(t_n)  - \sqrt{ \kappaC} \psi_{10g}^{\scriptscriptstyle (1)}(t_m, t_n) \bigg) ~\\
   + \sqrt{2}\xi_{\rm{in}}(t_m) \Big( \xi_{\rm{in}}(t_n) - \sqrt{ \kappaC}\psi_{10g}^{\scriptscriptstyle (1)}(t_m, t_n) \Big)   \bigg], \eqlab{in-out two app}
\end{multline}  
\end{subequations}
where $t_m\leq t_n$. The output state for two-photon inputs is 
\begin{align}\eqlab{output state two photons}
    \ket{\psi^{\scriptscriptstyle (2)}_{\rm{out}}} &\!= \!\!\int_{t_0}^{\Tend} \!\!\!\! \int_{t_0}^{\Tend} \!\!\!\!\!  dt_m dt_n \xi_{\rm{out}}^{\scriptscriptstyle (2)}(t_m, t_n)  \hat{w}^\dagger\hspace{-0.3mm}(t_m) \hat{w}^\dagger\hspace{-0.2mm}(t_n) \ket{\emptyset}\hspace{-0.3mm}.
\end{align}

The input-output relation for one-photon inputs are found by considering the two paths starting from the state $\ket{00g}\ket{1_k \bm{1}_m}$, which may be considered the single-photon branch of the map in~\figref{input-output map}. The result is 
%
\begin{align}\eqlab{input output one photon} 
    \xi_{\rm{out}}^{\scriptscriptstyle (1)}(t) &= \xi_{\rm{in}}(t) - \sqrt{\kappaC}\psi_{10g}^{\scriptscriptstyle (1)}(t),
\end{align}  
with a single-photon output state given by 
\begin{align}\eqlab{output state one photon}
    \ket{\psi^{\scriptscriptstyle (1)}_{\rm{out}}} &\!= \!\!\int_{t_0}^{\Tend} \!\!\!\!  dt \xi_{\rm{out}}^{\scriptscriptstyle (1)}(t) \hat{w}^\dagger(t) \ket{\emptyset} .
\end{align}

%% file: main_cphase_gate_application.tex
Having presented the equations governing the general time-evolution of an input state in product-form, we turn to the specific example of implementing a controlled-phase gate on two dual-rail encoded photonic qubits. Other quantum logic operations are in principle possible as well, but the controlled-phase gate is a prototypical example of a low-level two-qubit operation. Together with the available continuous single-qubit gates it completes the requirements for universal quantum circuits. \figref{PIC sketch} sketches the envisioned photonic integrated circuit implementation. The basic idea is that we arrange our TLE-cavity systems to act as an identity operation on incoming single-photon wavepackets (or the vacuum), while at the same time they impart a non-trivial phase to a two-photon wavepacket. The dual rail encoding and the beamsplitter ensure that the cavities encounter two-photon wavepackets only for the logical $\ket{11}$ state, leading to our controlled-phase operation.

\begin{figure}[h]
\centering
  \includegraphics[width=4.0cm] {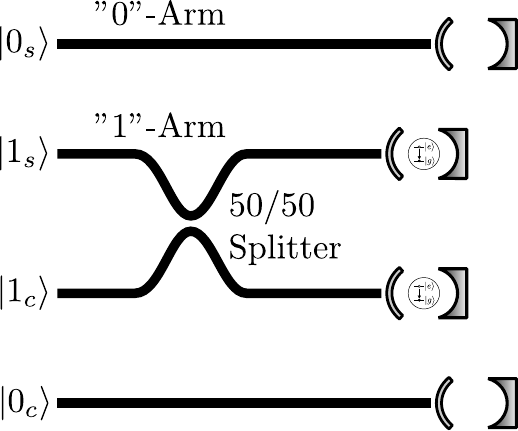} 
\caption{Photonic integrated circuit implementation of a controlled-phase gate on two dual-rail qubits encoded in four waveguides. The top two waveguides serve to encode one qubit. The logical state of that qubit depends on which of the waveguides contains a photon, as labeled on their left end. 
All four waveguides are terminated by identical one-sided cavities depicted in~\figref{concept}(a). The ``0''-arm cavities preserve the relative timing of the photon pulses and need not contain TLEs. The ``1''-arm cavities perform the nonlinear phase-shift and must contain identical TLEs. A 50/50 beam splitter between the ``1''-arms cause the transformation $\ket{1_w1_w}\rightarrow 1/\sqrt{2}(\ket{2_w0_w}+\ket{0_w2_w})$ resulting in two photons arriving at one of the TLE cavities if the logical state of the qubits is $\ket{11}$ (i.e., a photon in each of the middle two waveguides.). The cavity control is such that they map $\ket{1_w}\mapsto\ket{1_w}$, but $\ket{2_w}\mapsto-\ket{2_w}$. This extra phase is the crucial component enabling our controlled-phase gate. We use the $w$ subscript to denote physical photon in a waveguide, to avoid confusion with the notation used for the logical states of the dual-rail qubits.}
\figlab{PIC sketch}
\end{figure} 

The input state (two arbitrary dual-rail encoded qubits) is~\cite{Nysteen2017} 
\begin{multline}\eqlab{input qubit}
    \ket{\psi_{\rm{sc}}} = \big(\alpha\ket{0_s} + \beta\ket{1_s} \big) \otimes \big(\zeta\ket{0_c} + \vartheta\ket{1_c} \big) ~\equiv \\
    \alpha\zeta\ket{00} + \alpha\vartheta\ket{01} + \beta\zeta\ket{10} + \beta\vartheta\ket{11},
\end{multline}
with $|\alpha|^2+|\beta|^2=1$ and $|\zeta|^2+|\vartheta|^2=1$. The ideal controlled-phase gate operation is defined by the transformation
\begin{multline}\eqlab{c-phase gate operationt}
    \hat{\mathcal{C}} \ket{\psi_{\rm{sc}}} \equiv 
    \alpha\zeta\ket{00} + \alpha\vartheta\ket{01} + \beta\zeta\ket{10} - \beta\vartheta\ket{11}.
\end{multline}
We use the "worst-case" gate fidelity as defined in~\cite{Chuang2011}
\begin{align}
    \mathcal{F}_{\rm{gate}} \equiv \min_{\ket{\psi_{\rm{sc}}}} \!\big( \mathcal{F}_s \big),  
\end{align}
where the state fidelity, $\mathcal{F}_s$, is defined as 
\begin{multline}\eqlab{state fidelity qubit}
    \mathcal{F}_s \equiv \big|\langle \psi_{\rm{sc}} | \hat{\mathcal{C}}^\dagger |\psi_{\rm{out}}\rangle \big|^2 = \\
    \Big| \big( |\alpha|^2 \!+\! |\beta\zeta|^2\big ) \braket{1_w}{\psi_{\rm{out}}^{\scriptscriptstyle (1)}}^2
    - |\beta\vartheta|^2 \braket{2_w}{\psi_{\rm{out}}^{\scriptscriptstyle (2)}}  \Big|^2,
\end{multline}
where $\ket{1_w}$ and $\ket{2_w}$ denote, respectively, one or two photons in the waveguide and $\ket{\psi_{\rm{out}}^{\scriptscriptstyle (1)}}$ and $\ket{\psi_{\rm{out}}^{\scriptscriptstyle (2)}}$ denote the emitted state after the absorption of, respectively, one or two incident photons. We use the $w$ subscript to avoid confusion with the dual-rail logical states, like $\ket{1}=\ket{0_w1_w}$ and $\ket{0}=\ket{1_w0_w}$. The steps in the derivation of~\eqref{state fidelity qubit} are included in~\appref{app fid}. Notice the minus sign in the second term, corresponding to the fact that the logical $\ket{11}$ state has changed its phase, i.e., that when two photons are absorbed the state gains an additional $\pi$ phase, unlike when one or zero photons are absorbed.  
The complex-valued overlap factors in~\eqref{state fidelity qubit} are given by~\cite{Heuck2020a}
\begin{subequations}\eqlab{fidelity overlaps}
\begin{align}
\braket{1_w}{\psi_{\rm{out}}^{\scriptscriptstyle (1)}} &\equal \!\!\int \!\xi_{\rm{out}}^{\scriptscriptstyle (1)}(t) \xi_{\rm{in}}^{\hspace{-0.2mm}*}(t\minus \Tgate\hspace{0.2mm}) dt \\
    \braket{2_w}{\psi_{\rm{out}}^{\scriptscriptstyle (2)}} &\equal \!\!\int \!\!\!\int \!\!\xi_{\rm{out}}^{\scriptscriptstyle (2)}\hspace{-0.3mm}(\hspace{-0.3mm}t_m \hspace{-0.3mm} ,t_n\hspace{-0.3mm}) \xi_{\rm{in}}^{\hspace{-0.2mm}*}\hspace{-0.3mm}(\hspace{-0.3mm}t_n\!\!-\!\Tgate\hspace{0.2mm})\xi_{\rm{in}}^{\hspace{-0.2mm}*}\hspace{-0.3mm}(\hspace{-0.3mm}t_m\!\!-\!\Tgate\hspace{0.2mm})dt_ndt_m \hspace{-0.2mm} ,
\end{align}
\end{subequations}
where $T$ is the gate duration. Note that the output wave packet of the ideal gate operation is a simple time-translation of the input wave packet. This is a critical requirement for enabling quantum circuits with many identical gates, as any subsequent gate would work only if the wave packets carrying the encoded photons are not distorted by the previous gate.{{ The output wave packets described by $\xi_{\rm{out}}^{\scriptscriptstyle (1)}$ and $\xi_{\rm{out}}^{\scriptscriptstyle (2)}$ are not normalized due to loss and $|\psi_{00e}(t_{\scriptscriptstyle N})|^2 \geq 0$. The overlap integrals in~\eqref{fidelity overlaps} therefore describes gate errors in both amplitude and phase. }} \\

For the system considered here, the task is to determine the control fields $\Lambda(t)$ (the interaction between the cavity modes) and $\Omega(t)$ (the detuning between the TLE and cavity mode $\hat{b}$) that maximize the gate fidelity. Unity fidelity is achieved if $\braket{1_w}{\psi_{\rm{out}}^{\scriptscriptstyle (1)}}\equal 1$ and $\braket{2_w}{\psi_{\rm{out}}^{\scriptscriptstyle (2)}}\equal -1$ as seen from~\eqref{state fidelity qubit}. \emph{This means that a two-photon wave packet captured and then released by the cavity must acquire a different phase than that of a single photon to fulfill the condition $\arg[\xi_{\rm{out}}^{\scriptscriptstyle (2)}] - 2\arg[\xi_{\rm{out}}^{\scriptscriptstyle (1)}] \equal \pi$}. The photon-number dependent TLE-cavity coupling illustrated in~\figref{concept} causes an an-harmonic energy-ladder that enables this difference in phase-accumulation. However, in Refs.~\cite{Heuck2020, Heuck2020a} we found that the gate fidelity is limited due to interactions between the photons while the wave packet is absorbed and released from the cavity. This fidelity reduction would be particularly detrimental with the large nonlinearity considered here without a method to modify the effective size of the nonlinear coupling rate. Instead of changing $g$ itself, we consider modifying the TLE-cavity detuning, $\Omega(t)$. When $\Omega\!\gg\! g$, the effective nonlinearity is small and it is maximized when $\Omega\equal 0$. The gate protocol therefore consists of three stages: 
\begin{itemize}
\item[\emph{Absorption}:]{$\Lambda(t)$ is adjusted to couple photons from an incident wave packet into mode $\hat{b}$ while the detuning is held fixed at a large value $\Omega(t)\equal \Omega_0\!\gg\! g$.}
\item[\emph{Interaction}:]{$\Omega(t)$ is adjusted to increase the effective nonlinear coupling rate such that the required phase shift is achieved while the TLE returns to its ground state at the end of the stage for both one- and two-photon inputs.}
\item[\emph{Emission}:]{$\Lambda(t)$ is turned on again to release the photons into a wave packet with the same shape as the input while $\Omega(t)\equal \Omega_0$.}
\end{itemize}
When the TLE and cavity are completely decoupled, the optimum control function that loads a single photon into mode $\hat{b}$ is ~\cite{Heuck2020}
\begin{subequations}\eqlab{Lambda sol L main chi2} 
\begin{align}
    |\Lambda_i(t)|   &= \frac{|f_i| \exp[-\frac{\kappaL t}{2}] }{|\xi_{\rm{in}}|\sqrt{ 2\!\int_{t_0}^t\!f_i(s)ds } }\\
    \arg[\Lambda_i(t)]    &= -\delta_b t - \arg(\xi_{\rm{in}}),  \\
   f_i(t)         &= \Big(\frac{\kappaC}{2} \xi_{\rm{in}} - \dot{\xi}_{\rm{in}}\Big)\xi_{\rm{in}}^{*} e^{\kappaL t} \eqlab{fi definition main},
\end{align} 
\end{subequations}
where $\delta_b\equal 0$ and we assumed $\Lambda_i(t)$ arises due to three-wave mixing between modes $\hat{a}$, $\hat{b}$, and a third mode [not shown in~\figref{concept}(a)] occupied by a strong classical laser field. In the limit $\Omega_0\!\gg\! g$, we can adiabatically eliminate $\psi_{00e}^{\scriptscriptstyle (L)}$ from~\eqref{eoms psi01gM} by setting $\dot{\psi}_{00e}^{\scriptscriptstyle (L)}\approx 0$ in~\eqref{eoms psi00eM}, leading to 
\begin{subequations}\eqlab{eoms two modes two photons approx}
\begin{align}
    \dot{\psi}_{01g}^{\scriptscriptstyle (L)}( t_m ,t)  \approx &\Big( -\frac{ \kappaL}{2}  + i\frac{g^2}{\Omega_0} \Big) \psi_{01g}^{\scriptscriptstyle (L)}( t_m ,t) ~\nn \\
    &- i\Lambda(t)\psi_{10g}^{\scriptscriptstyle (L)}( t_m ,t) \eqlab{eoms psi01gM approx}\\
    \dot{\psi}_{00e}^{\scriptscriptstyle (L)}( t_m ,t)  \approx &-\frac{g}{\Omega_0} \psi_{01g}^{\scriptscriptstyle (L)}( t_m ,t) \eqlab{eoms psi00eM approx}.
\end{align} 
\end{subequations}
The term $g^2/\Omega_0$ therefore corresponds to adding an effective detuning in~\eqref{eoms psi01gM} so we add $g^2/\Omega_0 t$ to the phase of $\Lambda_i(t)$ when solving for the full dynamics described by~\eqref{eoms two modes two photons}. An alternative derivation of this additional phase term is found in~\appref{dressed state picture}.

The control function that optimally releases a single photon into the wave packet $\xi_{\rm{out}}$, is~\cite{Heuck2020}
\begin{subequations}\eqlab{Lambda sol out main chi2} 
\begin{align}
    |\Lambda_o(t)|  &= \frac{|f_o| e^{-\frac{\kappaL t}{2}}}{|\xi_{\rm{out}}|\sqrt{ \kappaC|\psi_{01}^{\scriptscriptstyle (1)}(t_0)|^2 \!-\! 2\!\int_{t_0}^t\!f_o(s)ds } } \eqlab{LAMo definition main}\\
    \arg\!\big[ \Lambda_o(t)\big]     &=  -\delta_b t - \arg(\xi_{\rm{out}}) -\frac{\pi}{2} \\
    f_o(t)     &= \Big(\frac{\kappaC}{2}\xi_{\rm{out}} + \dot{\xi}_{\rm{out}}\Big)\xi_{\rm{out}}^* e^{\kappaL t} \eqlab{fo definition main}.
\end{align} 
\end{subequations}
Note there is some additional optimization involved when $\psi_{01g}^{\scriptscriptstyle (1)}(0,t)$ has not reached a steady-state value at the onset of the release process since it is not obvious how to chose $\psi_{01g}^{\scriptscriptstyle (1)}(t_0)$ in~\eqref{LAMo definition main}. Since both $\Lambda_i$ and $\Lambda_o$ are approximately zero during the interaction stage, we have $\Lambda\equal \Lambda_i + \Lambda_o$.


%

%% file: main_gate_performance.tex
To quantify the gate performance that is possible with the system in Figs. \ref{fig:concept} and \ref{fig:PIC sketch}, we consider Gaussian-envelope wave packets
\begin{align}\eqlab{Gaussian main}
\xi_{\rm{in}}(t) &=  \!\sqrt{\frac{2}{\tauG}} \!\left(\frac{\text{ln}(2)}{\pi}\right)^{\!\!\frac{1}{4}} \!\!\exp\!\left(\!-2\text{ln}(2)\frac{(t-T_{\rm{in}})^2}{\tauG^2} \right), 
\end{align} 
where $|\xi_{\rm{in}}(t)|^2$ has a full temporal width at half maximum (FWHM) of $\tauG$ and a spectral width of $\OmG\equal 4\text{ln}(2)/\tauG$. We numerically solved the equations of motion in~\secref{hej} using \texttt{Julia}~\cite{bezanson2017julia}. The temporal shape of the control field $\Omega(t)$ was determined by minimizing the gate error $1\!-\!\Fgate$ using a standard gradient-free optimization method (Nelder-Mead~\cite{Mogensen2018}).~\figref{gate dynamics example} shows an example of the gate dynamics for a duration of $\Tgate\equal 7/g$, $T_{\rm{in}}\equal 4.3/g$, and $g\equal 0.4\OmG$. 
\begin{figure}[!h]  
\centering
  \includegraphics[width=8.0cm] {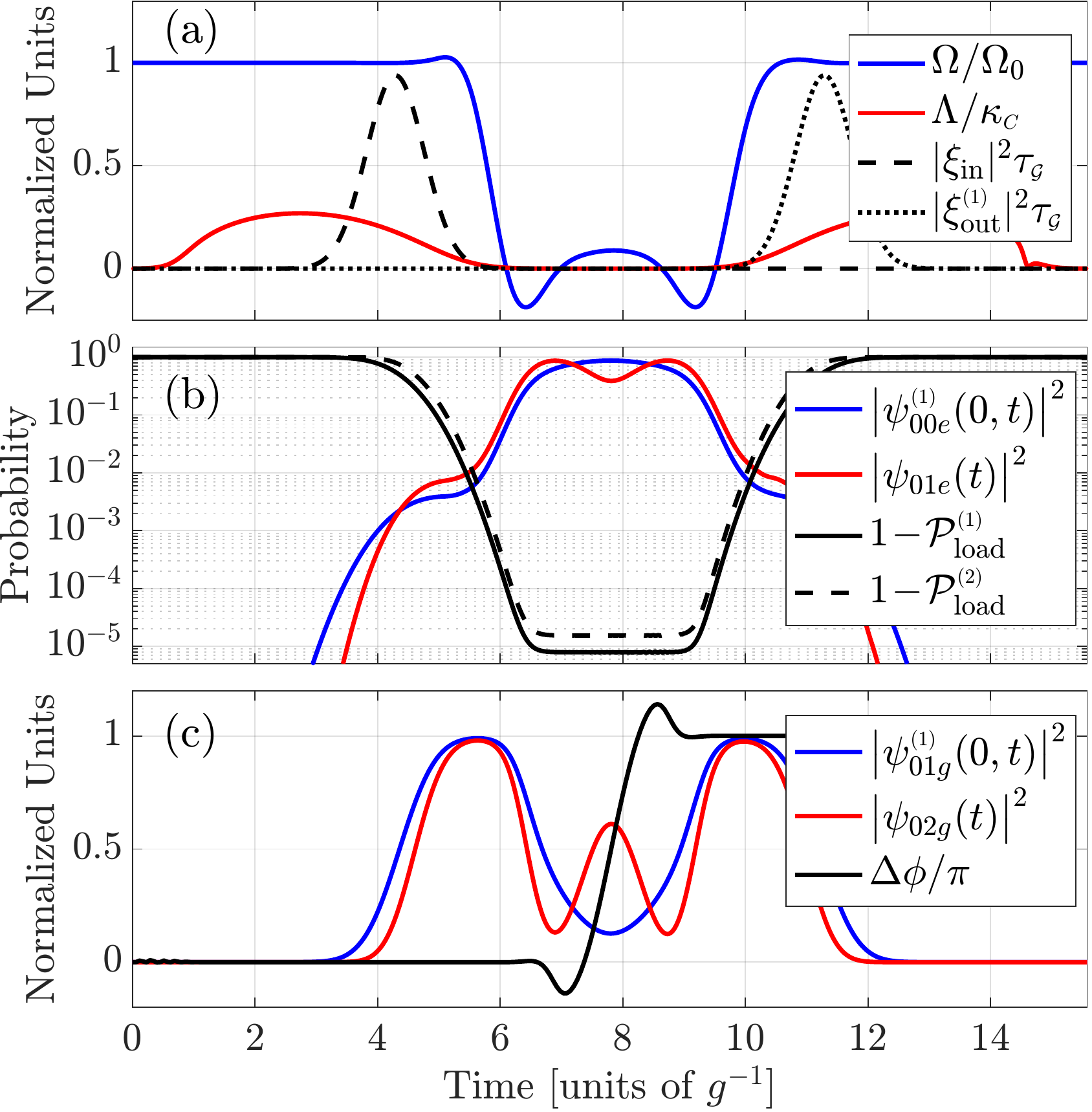} 
\caption{Example of gate dynamics. (a) Control functions, $\Lambda(t)$ and $\Omega(t)$ as a function of time along with the input wave packet and ideal one-photon output wave packet, $\xi_{\rm{out}}^{\scriptscriptstyle (1)}$. (b) Probability of the TLE being excited for a one-photon input, $\psi_{00e}^{\scriptscriptstyle (1)}(0,t)$, and the probability of a photon in mode $\hat{b}$ and an excited TLE for a two-photon input, $\psi_{01e}(t)$. The black curves plot the probability of having absorbed all the input photons [defined in~\eqref{Pload main}] for a one- (solid) and two-photon input state (dashed). (c) Probability that all input photons are in mode $\hat{b}$ for a one- (blue) and two-photon input state (red) along with the phase difference between the amplitudes of the corresponding Scr\"odinger coefficients [defined in~\eqref{phase diff main}] (black). Simulation parameters: $\kappaC\equal 6\OmG$, $\kappaL\equal \gamma_e\equal 0$, $g\equal 0.4\OmG$, $\Omega_0\equal 15g$, $T_{\rm{in}}\equal 4.3/g$, and $T\equal 7/g$.}
\figlab{gate dynamics example}
\end{figure} 
It is expected that the TLE-cavity detuning becomes small during the \emph{interaction} stage, $t\in[ T_{\rm{in}}; ~T_{\rm{in}}\!+\!\Tgate]$, since it leads to a larger occupation probability of the TLE and thereby a larger effective nonlinearity. The blue curve in~\figref{gate dynamics example}(a) confirms this expectation and~\figref{gate dynamics example}(b) plots the probability of the TLE being in the excited state for both one- (blue) and two-photon (red) input states. Note that both populations decrease towards zero at the end of the gate sequence as is required for a large gate fidelity. While the TLE-cavity detuning is low, the one- and two-photon states acquire phase at different rates, which is discussed in more detail in~\appref{dressed state picture}.~\figref{gate dynamics example}(c) plots the phase difference
\begin{align}\eqlab{phase diff main}
\Delta \phi (t) \equiv  \arg\!\big[ \psi_{02g}(t)\big] - 2\arg\!\big[ \psi_{01g}^{\scriptscriptstyle (1)} \hspace{-0.2mm} (0,t)\big],
\end{align} 
which approximates the phase difference between the output wave packets, $\arg[\xi_{\rm{out}}^{\scriptscriptstyle (2)}] - 2\arg[\xi_{\rm{out}}^{\scriptscriptstyle (1)}]$. The reason is that the populations $|\psi_{02g}(t)|^2$ and $|\psi_{01g}^{\scriptscriptstyle (1)} \hspace{-0.2mm} (0,t)|^2$ approach one immediately before the \emph{emission} stage as seen from~\figref{gate dynamics example}(c). 

\subsection{Absorption Efficiency}
The limitation on gate fidelity imposed by a finite value of $\Omega_0/g$ is observed in~\figref{gate dynamics example}(b) as a finite absorption probability (black lines). In~\figref{absorption}, we investigate this further by plotting the probability of not absorbing a one- or two-photon input state as a function of $\Omega_0/g$. 
\begin{figure}[h!]  
\centering
  \includegraphics[width=8.0cm] {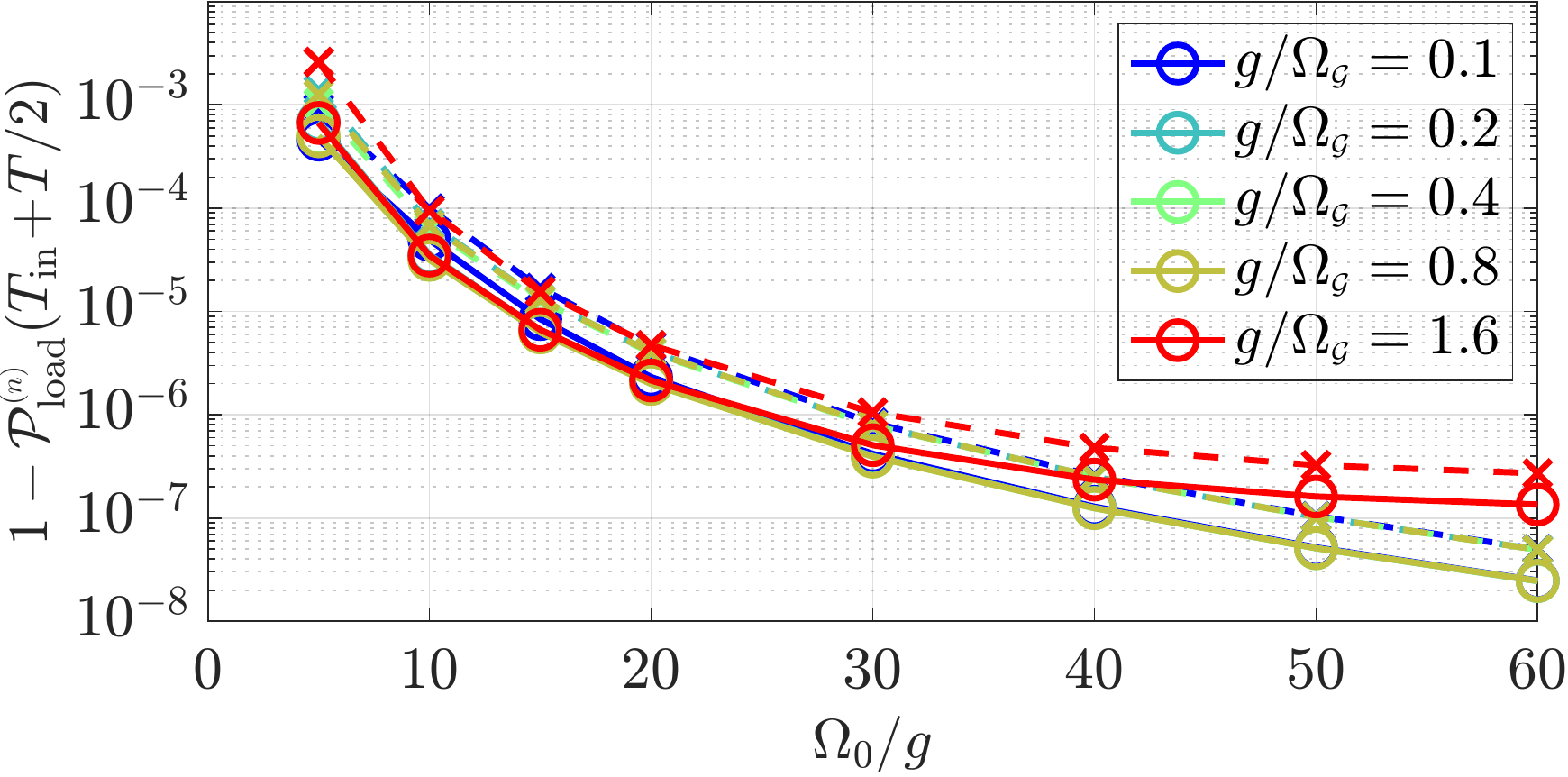} 
\caption{Probability of not absorbing one- (solid lines) and two-photon (dashed lines) input states as a function the TLE-cavity detuning for different values of $g/\OmG$. Simulation parameters: $\kappaC\equal 6\OmG$, $\kappaL\equal \gamma_e\equal 0$, $T_{\rm{in}}\equal 4.3/g$, and $T\equal 7/g$. }
\figlab{absorption}
\end{figure} 
The probabilities are given by 
\begin{subequations}\eqlab{Pload main}
\begin{align}
    \mathcal{P}_{\rm{load}}^{\scriptscriptstyle (1)}(t) &= \big| \psi_{01g}^{\scriptscriptstyle (1)} \hspace{-0.2mm} \big(0,t\big)\big|^2 + \big|\psi_{00e}^{\scriptscriptstyle (1)} \hspace{-0.2mm} \big(0,t\big)\big|^2 \\
    \mathcal{P}_{\rm{load}}^{\scriptscriptstyle (2)}(t) &= \big| \psi_{02g}\big(t\big)\big|^2 + \big|\psi_{01e}\big(t\big) \big|^2.
\end{align} 
\end{subequations}
The solution for the phase of the control function, $\Lambda(t)$, uses the term $g^2/\Omega_0 t$ derived in~\eqref{eoms two modes two photons approx} based on the approximation $\Omega_0\gg g$.~\figref{absorption} shows how the error probability increases as this approximation becomes worse for decreasing $\Omega_0/g$. Remarkably, the error for both one- and two-photon input states decreases rapidly with increasing $\Omega_0/g$ and drops to about $10^{-5}$ for $\Omega_0\equal 15 g$.\\


\subsection{Loss}
Our model includes a finite lifetime of cavity modes $\hat{a}$ and $\hat{b}$ as well as a decay rate from the TLE into the electromagnetic environment.~\figref{loss and dephasing}(a) plots the gate error as a function of gate duration for different values of the loss rate, $\kappaL$. Note that we assumed $\gamma_e\equal \kappaL$ in~\figref{loss and dephasing}(a). 
The control function, $\Omega(t)$, was optimized for each parameter configuration. The black line in~\figref{loss and dephasing}(a) sets a lower limit on the gate error due to a finite excitation probability of the TLE at $\Tend$ as well as a finite absorption error, $1\!-\!\mathcal{P}_{\rm{load}}$. 
\begin{figure}[!h] 
\centering
  \includegraphics[height=4.0cm] {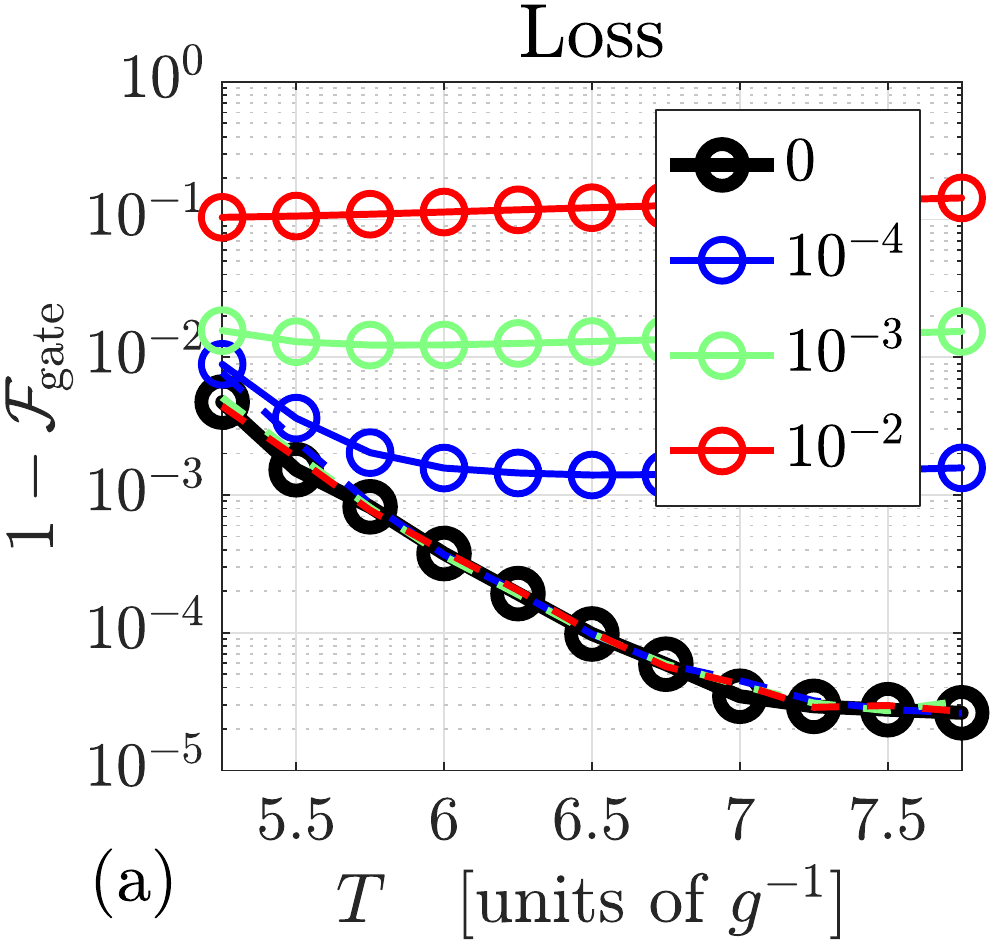} 
  \includegraphics[height=4.0cm] {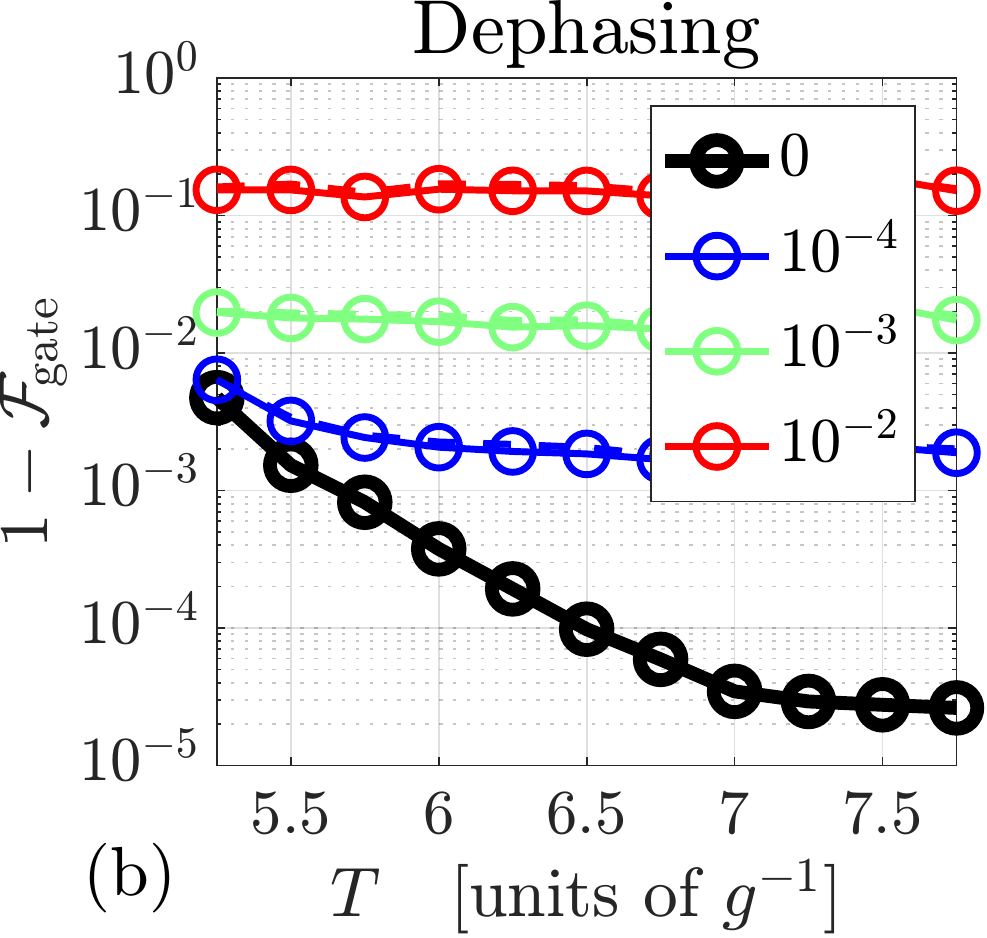} 
\caption{Dependence of the gate error on two types of decoherence. (a) $1-\Fgate$ as a function of gate duration, $\Tgate$, for different cavity loss rates (legend corresponds to $\kappaL/g$). (b) $1-\Fgate$ as a function of gate duration, $\Tgate$, for different dephasing rates (legend corresponds to $\gammaDP/g$). In both (a) and (b), the dashed lines plot the error in conditional fidelity. Simulation parameters: {{$\kappaC\equal 6\OmG$, $g\equal 0.4\OmG$, $\Omega_0\equal 15g$, and $T_{\rm{in}}\equal 4.3/g$. In (a) we used $\gammaDP\equal 0$ and $\gamma_e \equal \kappaL\neq 0$ and in (b) we used $\gamma_e \equal \kappaL \equal 0$ and $\gammaDP\neq 0$.}} }
\figlab{loss and dephasing}
\end{figure} 
Compared to Ref.~\cite{Heuck2020}, our analysis here studies all three stages of the gate sequence and the gate duration is more than three times shorter (when comparing~\figref{loss and dephasing}(a) here to figure 9 in~\cite{Heuck2020}). The dashed lines in~\figref{loss and dephasing}(a) correspond to the conditional fidelity~\cite{Heuck2020, Heuck2020a}, which is calculated using normalized output states
\begin{align}\eqlab{conditional output states}
    \ket{\overline{\psi}_{\rm{out}}^{\scriptscriptstyle (n)}} &\equiv \frac{ \ket{\psi_{\rm{out}}^{\scriptscriptstyle (n)}}}{\sqrt{\braket{\psi_{\rm{out}}^{\scriptscriptstyle (n)}}{\psi_{\rm{out}}^{\scriptscriptstyle (n)}}}}, ~~ n=\{1,2\}.
\end{align} 
It therefore corresponds to a post-selected gate fidelity conditioned on both photons being detected by a perfect detector. As expected, the conditional fidelity coincides with the fidelity in the absence of loss as in the case of $\chi^{\scriptscriptstyle (2)}$ and $\chi^{\scriptscriptstyle (3)}$ nonlinearities~\cite{Heuck2020a}.

Introducing control of the TLE-cavity detuning, $\Omega(t)$, removes the requirement observed in Ref.~\cite{Heuck2020a} to increase the gate duration, $\Tgate$, relative to the wave packet width, $\tauG$, in order to decrease the gate error due to wave packet distortions. Instead, $\Omega(t)$ controls the effective nonlinear coupling and the gate error (in the absence of loss) is only limited by the off-state detuning, $\Omega_0/g$, and the efficiency of depopulating the TLE for both one- and two-photon inputs despite the difference in Rabi frequency.

\subsection{Two-Level Emitter Dephasing }
Working with solid state quantum emitters introduces other types of error mechanisms in addition to loss. Energy-conserving interactions between the emitter and its environment may lead to dephasing, which means the coherence between the ground and excited state is lost~\cite{Lodahl2015}. Superposition states, $\alpha\ket{g} + \beta\ket{e}$, turn into mixed states when the relative phase between $\alpha$ and $\beta$ is not conserved. Here, we study this effect by introducing a dephasing rate, $\gammaDP$, and perform Monte-Carlo simulations to calculate the fidelity as described in~\appref{dephasing}.~\figref{loss and dephasing}(b) plots the gate error as a function of gate duration for different values of $\gammaDP$ while keeping $\kappaL\equal \gamma_e\equal 0$. 


The result is very similar to that in~\figref{loss and dephasing}(a), except the dashed and solid lines coincide in~\figref{loss and dephasing}(b). Dephasing errors can therefore be considered more severe than loss errors because the post-selected gate fidelity is also affected by dephasing.

%% file: main_noise_in_control_fields.tex
In this section, we consider a particular experimental approach to synthesizing the control fields and investigate the effect of noise in the settings of control parameters for $\Omega(t)$. A detuning between the emitter and cavity mode $\hat{b}$ could be controlled via the emitter transition energy, $\omega_e$, through AC-Stark shifts. An alternative scheme would be to modulate the cavity resonance, $\omega_b$, via e.g. cross-phase modulation. For experimentally demonstrated nonlinear coupling rates of $g \!\sim\! 40\,$GHz~\cite{Ota2018}, the entire gate duration in~\figref{gate dynamics example} is $T\!\sim\!175\,$ps, which would require very fast electronics. On the other hand, femtosecond-scale resolution in shaping of optical pulses was demonstrated~\cite{Supradeepa2008}. Typically, optical pulse shaping is achieved by modifying a finite number of Fourier components of pulses using gratings and spatial light modulators~\cite{Weiner2011}. To emulate this process, we write the control field as a sum of super-Gaussians with complex amplitudes in the Fourier domain    
\begin{align}\eqlab{Fourier Control Field}
\tilde{\Omega}(\omega) = \Omega_0 \delta(\omega) - e^{-i\omega T_{\Omega}}\!\!\sum_{m=-\mathcal{N}}^{\mathcal{N}} \!\!\! \tilde{\Omega}^{(m)} e^{-\big(\frac{\omega - m\Omega_{\rm{ch}}}{\Omega_{\rm{ch}}} \big)^{\!6}} ,
\end{align} 
where $\delta(\omega)$ is the Dirac-delta distribution, $\mathcal{N} \equal (N_{\rm{ch}}-1)/2$, and $T_{\Omega}$ shifts $\Omega(t)$ on the time-axis. The number of Fourier components is $N_{\rm{ch}}$ each having a bandwidth of $\Omega_{\rm{ch}}$. Since $\Omega(t)$ is real-valued, the optimization consists of determining $T_{\Omega}$ along with $\tilde{\Omega}_R^{(m)} \equal {\rm{Re}}\{\tilde{\Omega}^{(m)}\}$ and $\tilde{\Omega}_I^{(m)} \equal {\rm{Im}}\{\tilde{\Omega}^{(m)}\}$ under the constraints $\tilde{\Omega}_R^{(m)} \equal \tilde{\Omega}_R^{(-m)}$ and $\tilde{\Omega}_I^{(m)} \equal -\tilde{\Omega}_I^{(-m)}$.~\figref{control Fourier} shows an example of an optimized control pulse that results in a gate performance similar to the control pulse in~\figref{gate dynamics example}a. To see how the gate performance is affected by the number of Fourier components and the channel bandwidth, we plot the minimized gate error as a function of $N_{\rm{ch}}$ and $\Omega_{\rm{ch}}$ in~\figref{control Fourier noise}a.    
\begin{figure}
\centering
  \includegraphics[width=4cm] {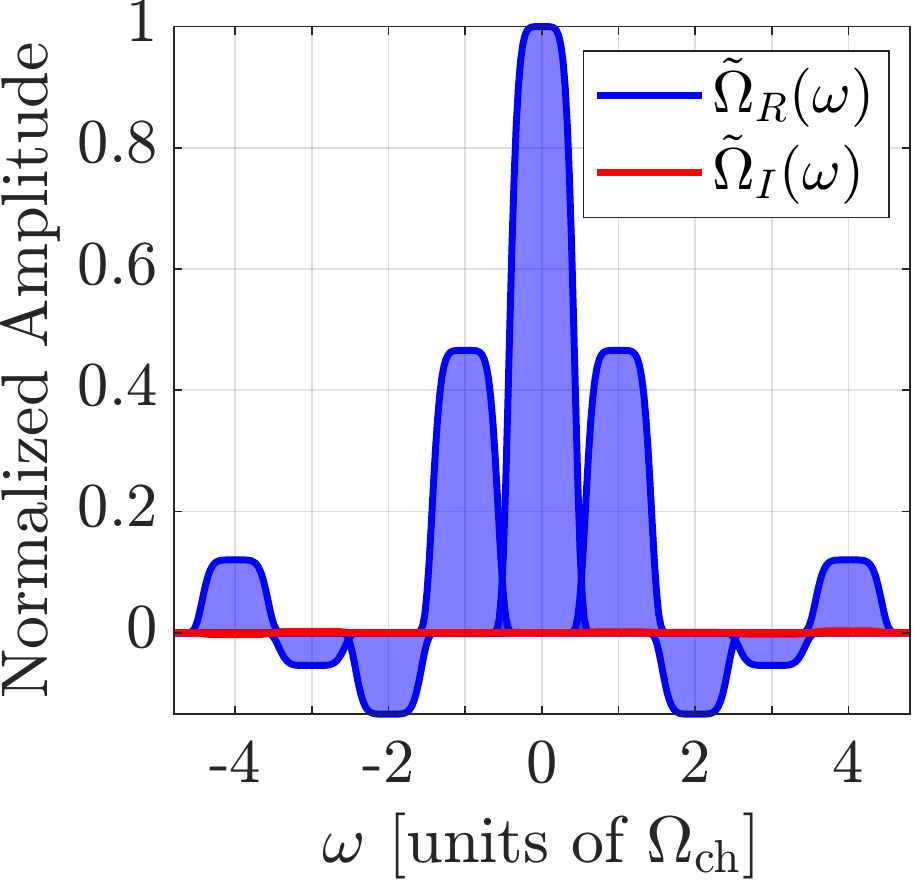} 
  \hspace{2mm}
  \includegraphics[width=4cm] {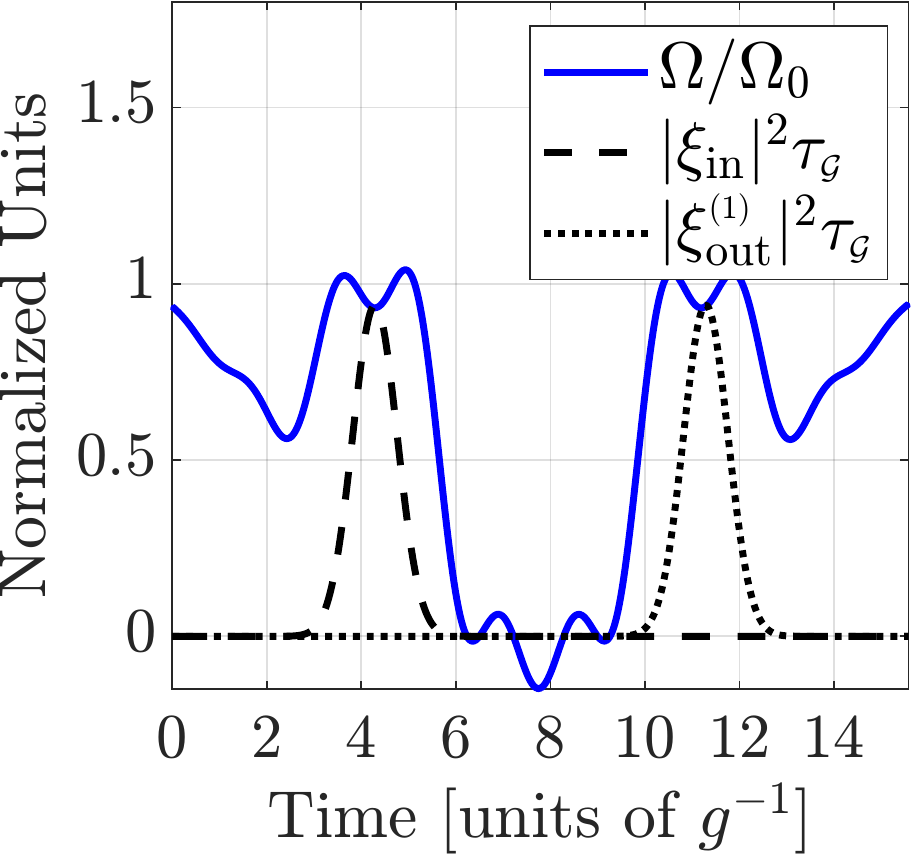} 
\caption{Example of Fourier domain control function leading to a gate error of $1-\Fgate\equal 4.7\!\times\!10^{-5}$.  Parameters: $\kappaC\equal 6\OmG$, $\gamma_e \equal \kappaL\equal 0$, $g\equal 0.4\OmG$, $\Omega_0\equal 15g$, $T\equal 7/g$, $T_{\rm{in}}\equal 4.3/g$, $N_{\rm{ch}}\equal 9$, and $\Omega_{\rm{ch}}/g \equal 0.14$. }
\figlab{control Fourier}
\end{figure} 

Experimentally, there is only a finite precision available to determine the shape of the control fields. The Fourier domain implementation enables a direct quantification of the effect on the gate error from noise in the complex amplitudes, $\tilde{\Omega}^{(m)}$, of a programmable filter. The noise is included by modifying the optimized real and imaginary control variables as
\begin{align}\eqlab{Noisy Fourier Control Field}
\tilde{\Omega}_{R/I}^{(m)} ~\rightarrow ~ \tilde{\Omega}_{R/I}^{(m)}  + X_{R/I}^{(m)} \!\times\!\sigma \!\times\! \max_{m}\!\big(\tilde{\Omega}_R^{(m)}, \tilde{\Omega}_I^{(m)}\big) .
\end{align} 
The size of the noise is represented by $\sigma$, $X_{R/I}^{(m)}$ is a random number between -1 and 1, and the last factor in~\eqref{Noisy Fourier Control Field} is the maximum of all the optimized variables. 
\begin{figure}[h!]  
\centering
  \includegraphics[height=3.7cm] {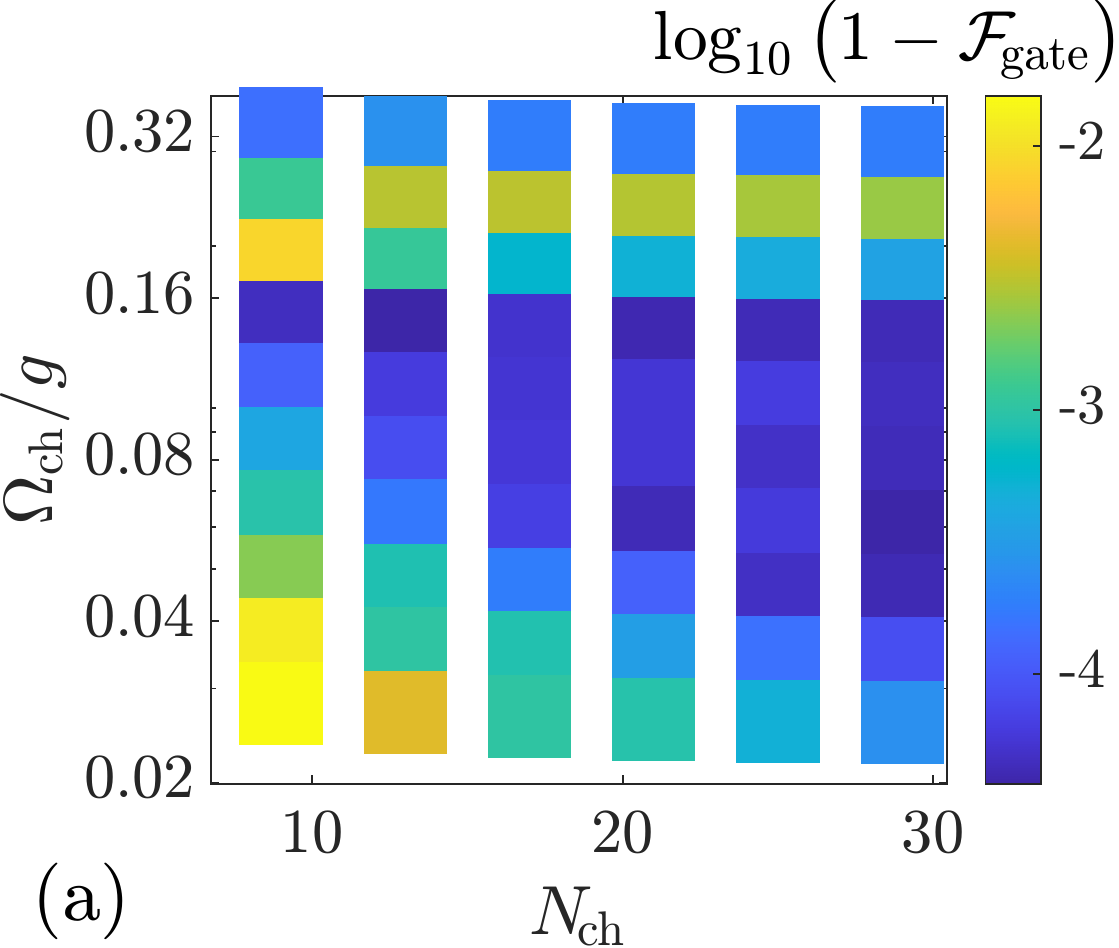} 
  \hspace{1mm}
  \includegraphics[height=3.48cm] {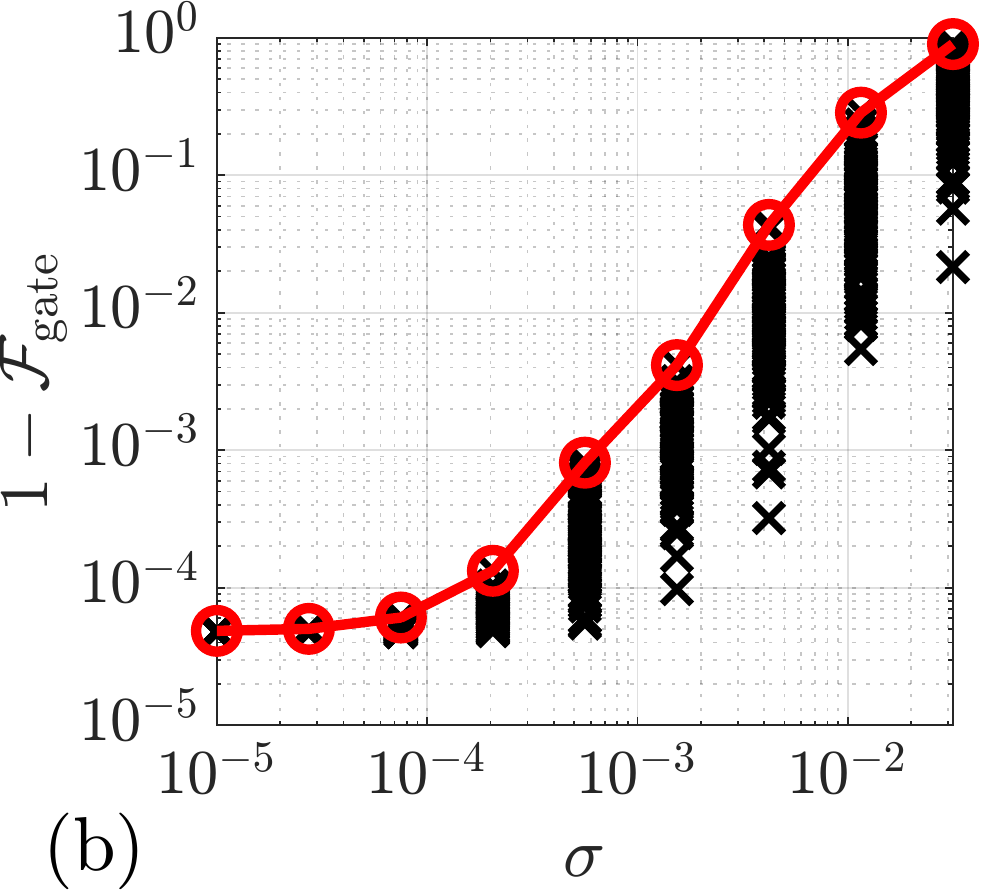} 
\caption{(a) Optimized gate error, $1-\Fgate$, as a function of the number of Fourier components, $N_{\rm{ch}}$, and their bandwidth, $\Omega_{\rm{ch}}$. (b) Gate error as a function of the noise parameter, $\sigma$, using 100 different combinations of random numbers, $X_{R/I}^{(m)}$, (black crosses). The red line shows the worst case scenario. Parameters: $\kappaC\equal 6\OmG$, $\gamma_e \equal \kappaL\equal 0$, $g\equal 0.4\OmG$, $\Omega_0\equal 15g$, and $T_{\rm{in}}\equal 4.3/g$. }
\figlab{control Fourier noise}
\end{figure} 
Using the maximum in~\eqref{Noisy Fourier Control Field} is motivated by a finite filter setting precision and represents an absolute error rather than a relative error. Adding noise degrades the gate fidelity and~\figref{control Fourier noise}(b) plots the gate error as a function of $\sigma$ using the same optimized parameters as in~\figref{control Fourier}. It is observed that errors below $10^{-4}$ are required to have a negligible influence on the gate error.

%% file: main_nonlinearity_comparison.tex
{{The nonlinearity required to facilitate photon-photon interactions for deterministic quantum logic gates can have different origins. In Refs.~\cite{Heuck2020, Heuck2020a}, we proposed protocols based on bulk nonlinearities such as second-harmonic generation (SHG) or self-phase modulation (SPM) in $\chi^{(2)}$ and $\chi^{(3)}$ materials.  By introducing a generalized nonlinear coupling rate, $\chiNL$, we may write the Hamiltonian describing three different nonlinear effects as~\cite{Heuck2020, Heuck2020a}
\begin{subequations}\eqlab{Hamiltonian nonlinear effects}
\begin{align}
    \hat{H}_{\rm{SHG}}   &\!=\! \hbar\chiNL \Big( \hat{c}\hat{b}^\dagger\hat{b}^\dagger + \hat{c}^\dagger\hat{b}\hat{b}\Big) \eqlab{H nonlinear chi2} \\
    \hat{H}_{\rm{SPM}}   &\!=\! \hbar \chiNL \Big(\hat{b}^\dagger\hat{b}\minus 1\Big)\hat{b}^\dagger\hat{b} \eqlab{H nonlinear chi3} \\
    \hat{H}_{\rm{TLE}}   &\!=\! \hbar \chiNL \Big(\hat{b}^\dagger \hat{\sigma}_{-}  + \hat{b} \hat{\sigma}_{+} \Big). \eqlab{H nonlinear TLE}
\end{align}
\end{subequations}
In~\eqref{H nonlinear chi2}, $\chiNL\!\propto\! \chill$, in~\eqref{H nonlinear chi3}, $\chiNL\!\propto\! \chilll$, and in~\eqref{H nonlinear TLE}, $\chiNL \equal g$ as seen from~\eqref{interaction picture Hamiltonian}. Note that we absorbed a factor of 1/4 into the definition of $\chiNL$ in~\eqref{H nonlinear chi3} compared to the definition of the $\chi_3$-parameter in equation 2b of Ref.~\cite{Heuck2020} to avoid any numerical pre-factors in~\eqref{Hamiltonian nonlinear effects}. Finding the minimum gate error for each value of $\kappaL$ in~\figref{loss and dephasing}(a) and plotting it as a function of $\chiNL/\kappaL$ shows that the error is approximately inversely proportional to $\chiNL/\kappaL$, see~\figref{gate comparison}.   
\begin{figure}[!h]
\centering
  \includegraphics[width=8.0cm] {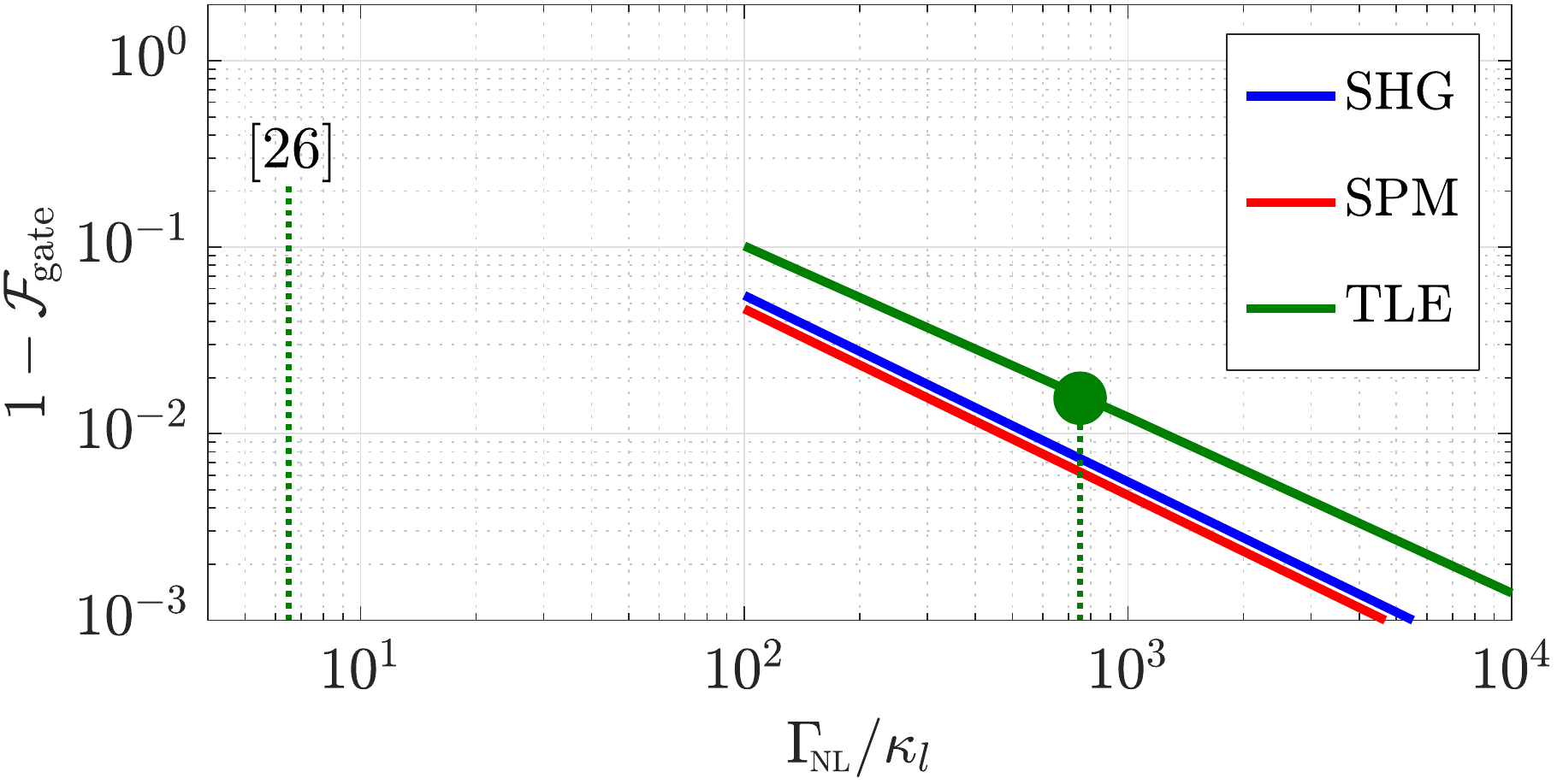} 
\caption{Gate error as a function of the ratio between nonlinear coupling rate and linear loss rate for the three types of nonlinearity we have studied here and in previous work. The generalized nonlinear coupling rate, $\chiNL$, corresponds to $g$ here and to $\chi_2$ or $\chi_3/4$ from equation 3 in Ref.~\cite{Heuck2020a}. The vertical dotted lines show state-of-the-art~\cite{Ota2018} and the expected performance by combining $g$ from Ref.~\cite{Ota2018} and $\kappaL$ from Ref.~\cite{Guha2017} (green dot). Note that the figure of merit, $\chiNL/\kappaL$, for bulk nonlinearities are orders of magnitude smaller than for the TLE and therefore not included in this plot. }
\figlab{gate comparison}
\end{figure} 
We also show the results from Ref.~\cite{Heuck2020} in the same plot for reference [note a rescaling of the red curve to match the definition of $\chiNL$ in~\eqref{H nonlinear chi3}]. A lower bound on the gate duration (in units of $\chiNL^{-1}$) follows from the physical origin of the phase difference between one- and two-photon inputs. For SHG, a full Rabi oscillation between two photons in mode $\hat{b}$ and one photon in mode $\hat{c}$ is required, and this Rabi period is given by $(\pi/\sqrt{2})\chiNL^{-1}$ (this can be seen from equations 57c and 57d in Ref.~\cite{Heuck2020a}). For SPM, the phase difference is simply acquired at a rate given by $2\chiNL t$ so the minimum required gate duration is $(\pi/2) \chiNL^{-1}$ (this can be seen from equation 54c in Ref.~\cite{Heuck2020a} when accounting for the definition: $\chiNL\equal\chi_3/4$). For TLEs, a bound is not as straightforwardly obtained due to the necessity of a control field to ensure the simultaneous achievement of a $\pi$ phase difference $\emph{and}$ depopulating the TLE for both one- and two-photon inputs. However,~\figref{loss and dephasing}(a) shows that a duration of $\sim \!5\chiNL^{-1}$ is sufficient for an error below 1\%. These bounds are consistent with the relative positions of the curves in~\figref{gate comparison} and shows that using TLEs as the optical nonlinearity comes with a relative small penalty in the required loss rate of a factor of 2-3 compared to $\chill$ or $\chilll$ effects.\\

To evaluate the potential of practical implementations, one must calculate the value of $\chiNL/\kappaL$.
%
\begin{table}[!t]
\renewcommand{\arraystretch}{1.4}
\vspace{3mm}
   \small
    \begin{tabular}{ |c|c|c|c|c|c|c|  }
    \hline
    \multirow{2}{*}{Ref.} &  \multirow{2}{*}{Mat.} &  \multirow{2}{*}{Type} & \multirow{2}{*}{${\displaystyle{\frac{\chiNL}{\rm{GHz}}}}$} & \multirow{2}{*}{$V_{\rm{int}}~\Big[{\displaystyle{\frac{\lambda^3}{n^3}\Big]}}$} & \multirow{2}{*}{$\QL$} & \multirow{2}{*}{$\displaystyle{\frac{\chiNL}{\kappaL}}$} \\
     & & & & & & \\
     \hline
     \multicolumn{7}{|c|}{Design Proposal} \\
     \hline
     \cite{Lin2016} & LiNbO$_3$ & $\chill$ & 0.01  & $1.1\!\times\!10^3$  & $2.4\!\times\!10^3$ & $1.2\!\times\!10^{-4}$ \\
     \hline
     \cite{Minkov2019} & GaN & $\chill$ & 0.00021  & $6.7\!\times\!10^4$  & $1\!\times\!10^4$ & $9.1\!\times\!10^{-6}$ \\
     \hline
     \cite{Choi2017} & Si & $\chilll$ & 0.002  & 0.17  & $2\!\times\!10^6$ & 0.02 \\
     \hline
     \multicolumn{7}{|c|}{Experimental Demonstration} \\
     \hline
     \cite{Lu2020} & LiNbO$_3$ & $\chill$ & 0.0012  & $7.4\!\times\!10^4$  & $5.8\!\times\!10^5$ & 0.0036 \\
     \hline
     \cite{Ota2018} & InAs & TLE & 40  & -  & $5.2\!\times\!10^4$ & 6.5 \\
     \hline
     \cite{Kuruma2020} & InAs & TLE & 4.8  & -  & $1.6\!\times\!10^5$ & 0.8 \\
     \hline
     \cite{Guha2017} & GaAs & - & -  & -  & $6.0\!\times\!10^6$ & - \\
     \hline
    \end{tabular}
    \caption{Comparison of nonlinear coupling rates and linear loss rates (given in terms of quality factors). For each material, we specify the type of the observed nonlinearity, the absolute value of the nonlinear interaction rate, the corresponding multi-mode interaction-volume, resonator quality factor, and the ration between the interaction rate and resonator decay rate. Definitions of interaction-volume for $\chill$ and $\chilll$ nonlinearity and their relation to other parameters listed in the literature are given in~\appref{mode volumes}.} 
    \tablab{comparison table}
\end{table}
\tabref{comparison table} lists numbers from the literature including each type of nonlinearity (see~\appref{mode volumes} for details on how the relevant metrics were extracted). It shows that interaction-volumes achieved for SHG in $\chill$-materials are orders of magnitude larger than those for both SPM in $\chilll$-materials and dipole interaction-volumes~\cite{Hu2016, Choi2017, Hu2018,Albrechtsen2021}. The dielectric confinement mechanism employed in Refs.~\cite{Hu2016, Choi2017, Hu2018,Albrechtsen2021} was applied to SHG in Ref.~\cite{Ateshian2021}, but more work is necessary to understand its potential for reducing the SHG interaction-volume. Using $\chiNL/\kappaL$ as a figure of merit is not generally applicable for SHG since two optical cavity modes are involved (note that we assumed identical loss rates for all modes in Ref.~\cite{Heuck2020a}). In the SHG literature, the conversion efficiency ($\eta_{\rm{SHG}}\!\propto\!Q_b^2Q_c$~\cite{Buckley2014,Lin2016, Minkov2019}) is often used as a figure of merit, but it appears that $\min(Q_b, Q_c)$ is the limiting factor in the quantum regime studied here. This difference in scaling of the figure of merit should be considered when designing cavities for few-photon interactions. 

High confinement cavities have been realized in Si~\cite{Hu2018,Albrechtsen2021}, but measurements of the SPM coupling rate are required to verify their potential.~\tabref{comparison table} clearly illustrates the advantage of two-level emitters compared to bulk nonlinearities in terms of the much larger nonlinear coupling rate.}} 

%% file: main_discussion.tex
{{In the introduction, we mentioned a few examples of two-level emitter implementations where strong coupling to an optical mode was already demonstrated. However, there has been a lot of work in recent years on other promising platforms like 1D-~\cite{He2018} and 2D-materials~\cite{Toth2019, Azzam2021}. Very strong coupling between excitons in 2D materials and plasmonic modes was also experimentally observed~\cite{Wen2017, Geisler2019, Qin2020} and theoretical work suggested how such systems may be described by an effective Jaynes-Cummings model~\cite{Tserkezis2020, Denning2020, Denning2020a}. Our focus on InAs quantum dots in GaAs membranes in the previous section and~\tabref{comparison table} is, however, based on our assessment that they represent state-of-the-art owing to their scalability potential and excellent properties resulting from a long history of developing them as single-photon sources. Moving beyond state-of-the-art and into a parameter regime corresponding to $\sim\!1\%$ gate error would require the nonlinear coupling rate in Ref.~\cite{Ota2018} $\emph{and}$ the linear loss rate in Ref.~\cite{Guha2017} to be achieved in the same device (illustrated with green dot in~\figref{gate comparison}). Surface passivation techniques are being used to address the challenge of achieving large $Q$s in GaAs cavities both with-~\cite{Kuruma2020} and without QDs~\cite{Guha2017}. Cavities with ultra-small dipole interaction-volumes~\cite{Hu2016, Choi2017, Hu2018,Albrechtsen2021, Ateshian2021} also represent an interesting approach to increase $g$. We note, however, that with the parameters used in~\figref{loss and dephasing} and $g\equal 40\,$GHz and $\omega_a\equal 2\pi c/940\,$nm~\cite{Ota2018}, the coupling-$Q$ of mode $\hat{a}$ is $Q_{\scriptscriptstyle C}\equal \omega_a/\kappaC\equal 530$. At the same time, $\kappaL/g\!\sim\!10^{-3}~\Rightarrow~Q_{l}/Q_{\scriptscriptstyle C}\equal 1.5\!\times\!10^4$ meaning that cavity mode $\hat{a}$ must be extremely over-coupled to reach the $\sim\!1\%$ gate error regime. Further increases to $g$ corresponds to an even smaller $Q_{\scriptscriptstyle C}$ that could pose experimental challenges although nanobeam cavities are well-suited to reach very over-coupled regimes even at large $Q$s~\cite{Quan2010}. 

The scheme for dynamic cavity coupling originating from nonlinear mode interactions used here and in recent work~\cite{Reddy2018, Heuck2019, Zhang2019, Heuck2020} is compatible with a very small dipole interaction-volume of the cavity mode interacting with the TLE. The control pump power may be increased to achieve the required strength of $\Lambda(t)$ as long as the overlap between the participating modes is large enough to ensure a reasonable nonlinear interaction-volume. However, interference-based dynamic cavity coupling~\cite{Tanaka2007,Xu2007} requires the mode to spread out across the interference paths and thereby limits how small the dipole interaction-volume can be.

In conclusion, we have shown that a two-level emitter is sufficient to implement high fidelity logical gates between photonic qubits when time-dependent control of the coupling between cavity modes and the emitter/cavity detuning is possible . 
Our approach represents a promising alternative to multi-level systems~\cite{Duan2004,Johne2012,Iakoupov2018} by shifting complexity from the atom-like emitter to the photonic system.

Based on the demonstrated performance and potential for improvement, we consider semiconductor quantum dots to be a very promising hardware platform to implement deterministic quantum logic on photonic quantum states.\\  }}


%% file: app_rotating_frame.tex
\section{Rotating Frame \applab{rotating frame}}
The Hamiltonian of the three cavity modes, pump field, and the TLE is $\hat{H}$, where
\begin{multline}\eqlab{Hamiltonian original frame app}
    \frac{\hat{H}}{\hbar} = \omega_a \hat{a}^\dagger\hat{a} +  \omega_b \hat{b}^\dagger\hat{b} + \omega_p \hat{p}^\dagger\hat{p} + \omega_e |e\rangle\langle e| + \omega_w\!\sum_{k=1}^N \hat{w}_k^\dagger\hat{w}_k ~+\\
    i \sqrt{\frac{\kappaC}{\Delta t}} \Big( \hat{a}^\dagger \hat{w}_n - \hat{a}\hat{w}_n^\dagger \Big)
       +  \chi_{\scriptscriptstyle \rm{DFG}} \Big(\hat{p}^\dagger \hat{a}^\dagger\hat{b} + \hat{p} \hat{b}^\dagger\hat{a}  \Big)  + g\big( \hat{b}^\dagger \hat{\sigma}_{-} + \hat{b}\hat{\sigma}_{+} \big) .
\end{multline} 
The operators are $\hat{\sigma}_{-}\!\equiv\! \ket{g}\bra{e}$, and $\hat{\sigma}_{+}\!\equiv\! \ket{e}\bra{g}$. The commutation relations are: $[\hat{a}^\dagger\hat{a}, \hat{a}] = -\hat{a}$ for the bosonic operators and $[\hat{\sigma}_{-}, |e\rangle\langle e|] \equal \hat{\sigma}_{-}$ and $[\hat{\sigma}_{+}, |e\rangle\langle e|] \equal -\hat{\sigma}_{+}$ for the fermionic operators. Notice that the commutators have the same structure if we substitute $\hat{a}^\dagger\hat{a},\ \hat{a},\  \hat{a}^\dagger$ for $|e\rangle\langle e|,\ \hat{\sigma}_{-},\ \hat{\sigma}_{+}$. We wish to move into the interaction picture, placing the evolution generated by the Hamiltonian $\hat{H}_0$ into the operators, where  
\begin{align}\eqlab{rotating frame unitary}
    \frac{\hat{H}_0}{\hbar} = \omega_{a}\hat{a}^\dagger\hat{a} + \omega_{b}\hat{b}^\dagger\hat{b}  + \omega_{b}|e\rangle\langle e| + \omega_p \hat{p}^\dagger\hat{p} +\omega_{w}\!\sum_{k=1}^N \hat{w}_k^\dagger\hat{w}_k . 
\end{align} 
Notice that we have taken the reference frequency of the TLE to be that of mode $\hat{b}$. We do so, as $\omega_e$ will actually be a time-dependent control parameter to be optimized and we would want to only occasionally bring the TLE in resonance with mode $\hat{b}$. Alternatively, the control could be applied to cavity mode $\hat{b}$ so that $\omega_b$ was effectively time-dependent instead of $\omega_e$. In that case, we would use $\omega_e$ in front of the term $\hat{b}^\dagger\hat{b}$ so that $\hat{H}_0$ would again be time-independent. The evolution of the state of the system is now given by an effective interaction Hamiltonian, usually referred to as the ``interaction Hamiltonian in the interaction picture", which is given by
\begin{align}\eqlab{Hamiltonian rotating frame definition}
    \hat{H}_{\rm{I}}(t) = \hat{U}\hat{H}\hat{U}^\dagger + i \hbar \hat{U} \frac{\partial \hat{U}^\dagger}{\partial t} = \hat{U}(\hat{H} - \hat{H}_0)\hat{U}^\dagger,
\end{align} 
where the second equality holds since $\hat{U} = e^{-i\hat{H}_0t/\hbar}$ and $\hat{H}_0$ has no time dependence. Let us evaluate~\eqref{Hamiltonian rotating frame definition} for terms in $\hat{H}$ that do not commute with $\hat{H}_0$.

We will use
\begin{align}\eqlab{exponential transform identity}
  e^{\alpha \hat{A}} \hat{B} e^{-\alpha \hat{A}} = \hat{B} + \alpha[\hat{A}, \hat{B}] + \frac{\alpha^2}{2!}\big[\hat{A}, [\hat{A}, \hat{B}]\big] + \ldots,
\end{align} 
which implies 
\begin{align}\eqlab{Hamiltonian rotating frame trans bosons 2}
 \hat{U} \hat{a} \hat{U}^\dagger = \hat{a}\sum_{n=0}^\infty \frac{1}{n!} \big(\!i\omega_{a} t\big)^n = \hat{a} e^{i\omega_a t}.
\end{align}

We perform the same calculation for $\hat{w}$ and $\hat{b}$, as well as their Hermitian conjugates. As we have seen that the corresponding commutators are the same for the TLE, we also get

\begin{align}\eqlab{Hamiltonian rotating frame trans bosons 2}
 \hat{U} \hat{\sigma}_{-} \hat{U}^\dagger = \hat{\sigma}_{-} e^{i\omega_b t}.
\end{align}

With this, we can evaluate~\eqref{Hamiltonian rotating frame definition} by inserting the appropriate exponential functions according to~\eqref{rotating frame unitary}
\begin{align}\eqlab{interaction picture Hamiltonian app}
    \frac{\hat{H}_{\rm{I}}(t)}{\hbar} =  \Omega(t)|e\rangle\langle e| + i \sqrt{\frac{\kappaC}{\Delta t}} \Big( \hat{a}^\dagger \hat{w}_n - \hat{a}\hat{w}_n^\dagger \Big) +
    \chi_{\scriptscriptstyle \rm{DFG}} \Big(\hat{p}^\dagger \hat{a}^\dagger\hat{b} e^{-i\delta_\Lambda t} + h.c.  \Big) + g\Big(\hat{b}^\dagger \hat{\sigma}_{-} + h.c. \Big).
\end{align} 
where we have taken
\begin{subequations}\eqlab{detunings TLE}
\begin{align}
         \qquad 0 &= \omega_a - \omega_w \\
     \Omega(t)  &\equiv \omega_e(t) - \omega_b \\
     \delta_\Lambda &\equiv \omega_p - (\omega_b-\omega_a) \equal 0.
\end{align} 
\end{subequations}

%% file: app_fidelity_equation.tex
\section{CPHASE Gate Fidelity\applab{app fid}}
In the main text the overlap between the desired target state and the actual obtained state was stated to be
\begin{equation}\eqlab{app state fidelity qubit}
    \mathcal{F}_s \equiv \big|\langle \psi_{\rm{sc}} | \hat{\mathcal{C}}^\dagger |\psi_{\rm{out}}\rangle \big|^2 =
    \Big| \big( |\alpha|^2 \!+\! |\beta\zeta|^2\big ) \braket{1_w}{\psi_{\rm{out}}^{\scriptscriptstyle (1)}}^2
    - |\beta\vartheta|^2 \braket{2_w}{\psi_{\rm{out}}^{\scriptscriptstyle (2)}}  \Big|^2
\end{equation}
where $\ket{1_w}$ and $\ket{2_w}$ are single- and two-photon wave packets propagating in the waveguide with shapes $\xi_{\rm{in}}(t)$ and $\xi_{\rm{in}}(t)\xi_{\rm{in}}(t')$, while  $\ket{\psi_{\rm{out}}^{\scriptscriptstyle (1)}}$ and $\ket{\psi_{\rm{out}}^{\scriptscriptstyle (2)}}$ are the actual states released into the waveguide according to our model. The possible logical states are the dual rail encoded states $\ket{00}$, $\ket{01}$, $\ket{10}$, and $\ket{11}$, where each of these two qubits can be written as $\ket{0}=\ket{1_w0_w}$ and $\ket{1}=\ket{0_w1_w}$, where the $w$ denotes that the given number of photons are stored in a waveguide, i.e., $\ket{1_w}=\int dt\xi_\mathrm{in}(t)\hat{w}^\dagger(t)\ket{\emptyset}$. In a slight abuse of notation we use $w$ without specifying in which of the four waveguides and at what point in time the wave packet is propagating. 

As described in the main text, our gate consist of the application of a beam splitter interaction ($\hat{B}$), followed by the active capture and release of photons ($\hat{C}$), followed by passing through the beam splitter in the opposite direction ($\hat{B}^\dagger$), i.e.,
\begin{equation}
    \hat{\mathcal{C}} = \hat{B}^\dagger\hat{C}\hat{B}.
\end{equation}

The initial beam splitter already moves the state of the system away from one that can be interpreted as dual-rail encoded qubits (in the following notation, the first two Fock numbers correspond to the two rails of the first qubit and the second two Fock numbers correspond to the two rails of the second qubit; the beam splitter connects the second rail of each aforementioned pair of rails):
\begin{subequations}
\begin{align}
        \hat{B}\ket{00} &= \hat{B}\ket{1_w0_w1_w0_w} = \ket{1_w0_w1_w0_w}, \\
        \hat{B}\ket{01} &= \hat{B}\ket{1_w0_w0_w1_w} = \frac{\ket{1_w0_w0_w1_w}+\ket{1_w1_w0_w0_w}}{\sqrt{2}}, \\
        \hat{B}\ket{10} &= \hat{B}\ket{0_w1_w1_w0_w} = \frac{-\ket{0_w1_w1_w0_w}+\ket{0_w0_w1_w1_w}}{\sqrt{2}}, \\
        \hat{B}\ket{11} &= \hat{B}\ket{0_w1_w0_w1_w} = \frac{\ket{0_w0_w0_w2_w}+\ket{0_w2_w0_w0_w}}{\sqrt{2}}.
\end{align}
\end{subequations}
After operating with $\hat{C}$, we have:
\begin{subequations}
\begin{align}
        \hat{C}\hat{B}\ket{00} &= \ket{\psi_{\rm{out}}^{\scriptscriptstyle (1)}0_w\psi_{\rm{out}}^{\scriptscriptstyle (1)}0_w}, \\
        \hat{C}\hat{B}\ket{01} &= \frac{\ket{\psi_{\rm{out}}^{\scriptscriptstyle (1)}0_w0_w\psi_{\rm{out}}^{\scriptscriptstyle (1)}}+\ket{\psi_{\rm{out}}^{\scriptscriptstyle (1)}\psi_{\rm{out}}^{\scriptscriptstyle (1)}0_w0_w}}{\sqrt{2}}, \\
        \hat{C}\hat{B}\ket{10} &= \frac{-\ket{0_w\psi_{\rm{out}}^{\scriptscriptstyle (1)}\psi_{\rm{out}}^{\scriptscriptstyle (1)}0_w}+\ket{0_w0_w\psi_{\rm{out}}^{\scriptscriptstyle (1)}\psi_{\rm{out}}^{\scriptscriptstyle (1)}}}{\sqrt{2}}, \\
        \hat{C}\hat{B}\ket{11} &= \frac{\ket{0_w0_w0_w\psi_{\rm{out}}^{\scriptscriptstyle (2)}}+\ket{0_w\psi_{\rm{out}}^{\scriptscriptstyle (2)}0_w0_w}}{\sqrt{2}}.
\end{align}
\end{subequations}
Notice that in the case of single photons the cavities containing a TLE emitter and the bare cavities are tuned to perform the same (identity) operation. Only the central two cavities ever see more than a single photon, and the nontrivial phase added to such a state is at the root of our c-phase gate. To find $\mathcal{F}_s=\big|\langle \psi_{\rm{out}} | \hat{B}^\dagger\hat{C}\hat{B} |\psi_{\rm{sc}}\rangle \big|^2$ we will calculate $\langle \psi_{\rm{out}} | \hat{B}^\dagger$ and $\hat{C}\hat{B} |\psi_{\rm{sc}}\rangle$ separately and then multiply them together, leading directly to the formula stated in the main text.

%% file: app_equations_of_motion.tex
\section{Occupation Probabilities \applab{app eoms}}

Derivations of the occupation probabilities follow the same procedure as in Appendix D of Ref.~\cite{Heuck2020}. Here,~\figref{input-output map} is used to keep track of all the terms contributing to the probabilities. The probability that both photons are in the waveguide, $\mathcal{P}_{00g}$, has contributions from one photon on the input side and one on the output side, both on the input side, and both photons on the output side. The first contribution is
\begin{multline}\eqlab{P_00g a} 
    \big|\expect{00g|00g}\big|^2 \!\sum_{j',k'=1}^N \bigg | \sum_{m=1}^n  \bigg ( \big[-\sqrt{\kappaC}\psi_{10g}^{\scriptscriptstyle (2)}(m) + \sqrt{2}\xi^{\rm{in}}_m \big] \sum_{k>m}  \xi^{\rm{in}}_k \Delta t   \expect{1_{j'} 1_{k'} | 1_k \mathbf{1}_m}  \bigg) \bigg|^2 ~=\\
    \sum_{k'=1}^{n}\sum_{j'=1}^{N} \bigg | \big[-\sqrt{\kappaC}\psi_{10g}^{\scriptscriptstyle (2)}(k') + \sqrt{2}\xi^{\rm{in}}_{k'} \big] \sum_{k>k'}  \xi^{\rm{in}}_k \Delta t   \expect{ 1_{j'} | 1_k }   \bigg|^2 ~=\\
    \sum_{k'=1}^{n} \!\Big|-\sqrt{\kappaC}\psi_{10g}^{\scriptscriptstyle (2)}(k') + \sqrt{2}\xi^{\rm{in}}_{k'} \Big|^2 \sum_{j'=1}^{N} \Big|\sum_{k>k'}  \Delta t \xi^{\rm{in}}_{j'}  \expect{ 1_{j'} | 1_k } \Big|^2 =
    \sum_{k'=1}^{n} \Delta t \Big|-\sqrt{\kappaC}\psi_{10g}^{\scriptscriptstyle (2)}(k') + \sqrt{2}\xi^{\rm{in}}_{k'} \Big|^2 \sum_{j'=k'}^{N} \Delta t\big | \xi^{\rm{in}}_{j'}  \big|^2,
\end{multline}
where the summation over $m$ from 1 to $n$ was included because the photon on the output side could be in any bin between 1 and $n$. The contribution from both photons being on the input side is
\begin{align}\eqlab{P_00g ab} 
    \big|\expect{00g|00g}\big|^2 \!\sum_{j',k'=1}^N \sum_{m'>n} \!\Delta t\big| \xi^{\rm{in}}_{m'}\big|^2 \sum_{m>n} \!\Delta t\big| \xi^{\rm{in}}_{m}\big|^2 \big|\expect{1_{j'} 1_{k'} | 1_{m'} 1_m} \big|^2  =\sum_{j'>n}\!\Delta t\big| \xi^{\rm{in}}_{j'}\big|^2 \sum_{k'>n} \!\Delta t\big| \xi^{\rm{in}}_{k'}\big|^2 .
\end{align}
Similarly, the contribution from the output state is
\begin{align}\eqlab{P_00g b} 
    \big|\expect{00g|00g}\big|^2 \!\sum_{j',k'=1}^N \sum_{m'=1}^n \sum_{m=1}^n \!\Delta t\Delta t\big| \xi^{\rm{out}}_{m'm}\big|^2 \big|\expect{1_{j'} 1_{k'} | \mathbf{1}_{m'} \mathbf{1}_m} \big|^2  =\sum_{m'=1}^n \sum_{m=1}^n \!\Delta t\Delta t\big| \xi^{\rm{out}}_{m'm}\big|^2 .
\end{align}
Adding the contributions from Eqs. (\ref{eq:P_00g a}) - (\ref{eq:P_00g b}) and taking the continuum limit, we get
\begin{multline}\eqlab{P_00g} 
    P_{00g}(t_n) =  \!\int_{t_0}^{t_n} \!\!\!\Big( \big|\sqrt{2}\xi_{\rm{in}}(t_m) - \sqrt{\kappaC}\psi_{10g}^{\scriptscriptstyle (2)}(t_m)\big|^2 \!\!\int_{t_m}^{t_{\!N}} \!\!\big|\xi^{\rm{in}}(s)\big|^2 ds \Big)dt_m ~+\\
    \bigg(\! \int_{t_n}^{t_{\!N}} \!\!\!|\xi_{\rm{in}}(t)|^2 dt \bigg)^{\!2} + \int_{t_0}^{t_n}\!\!\int_{t_0}^{t_n}\big|\xi_{\rm{out}}(t_m, s)\big|^2 dsdt_m .
\end{multline}
Similarly, we find the probability of one photon in the waveguide and the TLE in the excited state
\begin{multline}\eqlab{P_00e} 
    P_{00e}(t_n) = \big|\psi_{00e}^{\scriptscriptstyle (2)}(t_n)\big|^2 \!\int_{t_n}^{t_{N}} \!\!|\xi^{\rm{in}}(s)|^2 ds  ~+ \int_{t_0}^{t_n} \!\bigg| \Big[ \sqrt{2}\xi_{\rm{in}}(t_m) - \sqrt{\kappaC}\psi_{10g}^{\scriptscriptstyle (2)}(t_m) \Big] \psi_{00e}^{\scriptscriptstyle (1)}(t_m, t_n) ~-\\
    \sqrt{\kappaC} \Big[ \sqrt{2}\psi_{20g}(t_m)A_{00}(t_m, t_n) + \psi_{11g}(t_m)B_{00}(t_m, t_n) + \psi_{10e}(t_m)C_{00}(t_m, t_n) \Big] ~+\\
     \xi_{\rm{in}}(t_m) \Big[ \psi_{10g}^{\scriptscriptstyle (2)}(t_m) A_{00}(t_m, t_n) + \psi_{01g}^{\scriptscriptstyle (2)}(t_m)B_{00}(t_m, t_n) + \psi_{00e}^{\scriptscriptstyle (2)}(t_m)C_{00}(t_m, t_n)\Big] \bigg|^2 dt_m .
\end{multline}
The probability of one photon in the waveguide and one photon in mode $\hat{a}$ is
\begin{multline}\eqlab{P_10g} 
    P_{10g}(t_n) = \big|\psi_{10g}^{\scriptscriptstyle (2)}(t_n)\big|^2 \!\int_{t_n}^{t_{N}} \!\!|\xi^{\rm{in}}(s)|^2 ds  ~+ \int_{t_0}^{t_n} \!\bigg| \Big[ \sqrt{2}\xi_{\rm{in}}(t_m) - \sqrt{\kappaC}\psi_{10g}^{\scriptscriptstyle (2)}(t_m) \Big] \psi_{10g}^{\scriptscriptstyle (1)}(t_m, t_n) ~-\\
    \sqrt{\kappaC} \Big[ \sqrt{2}\psi_{20g}(t_m)A_{10}(t_m, t_n) + \psi_{11g}(t_m)B_{10}(t_m, t_n) + \psi_{10e}(t_m)C_{10}(t_m, t_n) \Big] ~+\\
     \xi_{\rm{in}}(t_m) \Big[ \psi_{10g}^{\scriptscriptstyle (2)}(t_m) A_{10}(t_m, t_n) + \psi_{01g}^{\scriptscriptstyle (2)}(t_m)B_{10}(t_m, t_n) + \psi_{00e}^{\scriptscriptstyle (2)}(t_m)C_{10}(t_m, t_n)\Big] \bigg|^2 dt_m .
\end{multline}
The probability of one photon in the waveguide and one photon in mode $\hat{b}$ is
\begin{multline}\eqlab{P_01g} 
    P_{01g}(t_n) = \big|\psi_{01g}^{\scriptscriptstyle (2)}(t_n)\big|^2 \!\int_{t_n}^{t_{N}} \!\!|\xi^{\rm{in}}(s)|^2 ds  ~+ \int_{t_0}^{t_n} \!\bigg| \Big[ \sqrt{2}\xi_{\rm{in}}(t_m) - \sqrt{\kappaC}\psi_{10g}^{\scriptscriptstyle (2)}(t_m) \Big] \psi_{01g}^{\scriptscriptstyle (1)}(t_m, t_n) ~-\\
    \sqrt{\kappaC} \Big[ \sqrt{2}\psi_{20g}(t_m)A_{01}(t_m, t_n) + \psi_{11g}(t_m)B_{01}(t_m, t_n) + \psi_{10e}(t_m)C_{01}(t_m, t_n) \Big] ~+\\
     \xi_{\rm{in}}(t_m) \Big[ \psi_{10g}^{\scriptscriptstyle (2)}(t_m) A_{01}(t_m, t_n) + \psi_{01g}^{\scriptscriptstyle (2)}(t_m)B_{01}(t_m, t_n) + \psi_{00e}^{\scriptscriptstyle (2)}(t_m)C_{01}(t_m, t_n)\Big] \bigg|^2 dt_m .
\end{multline}
The probabilities of both photons being in the system are easily found from the solutions to~\eqref{ODEs two photon terms}: $\mathcal{P}_{11g}\equal|\psi_{11g}|^2$, $\mathcal{P}_{20g}\equal|\psi_{20g}|^2$, $\mathcal{P}_{02g}\equal|\psi_{02g}|^2$, $\mathcal{P}_{10e}\equal|\psi_{10e}|^2$, and $\mathcal{P}_{01e}\equal|\psi_{01e}|^2$.

%% file: app_dressed_state_picture.tex
\section{Dressed State Picture \applab{dressed state picture}}
Let us consider the free evolution of cavity mode $\hat{b}$ and the TLE with a constant detuning, $\Omega(t)\equal \Omega_0$. We write the equations of motion in matrix form, $\ket{\dot{\bm{\psi}}} = \bm{A} \ket{\bm{\psi}}$. For a single photon, we have 
\begin{align}\eqlab{free evolution}
    \ket{\bm{\psi}} &=  \left[\begin{array}{c}  \psi_{01g} \\ \psi_{00e} \end{array} \right], ~\text{and} ~  \bm{A} = 
    \left[\begin{array}{cc}  -\frac{\kappaL}{2} \,\,\,\, & -i g \\ -ig\,\,\,\, & -\frac{\gamma_e}{2} - i\Omega_0 \end{array} \right].
\end{align}
%
The solution to the system of coupled first-order differential equations is 
\begin{align}\eqlab{solution free evolution}
   \left[\begin{array}{c}  \psi_{01g} \\ \psi_{00e} \end{array} \right] = 
    \left[\begin{array}{c}  v_{+,1} \\ v_{+,2} \end{array} \right] e^{\lambda_{+} t} + \left[\begin{array}{c}  v_{-,1} \\ v_{-,2} \end{array} \right] e^{\lambda_{-} t},
\end{align}
where ($\bm{v}_{+},\lambda_{+}$) and ($\bm{v}_{-},\lambda_{-}$) are the eigenvectors and eigenvalues of $\bm{A}$. They are given by
\begin{subequations}\eqlab{eigenvectors and eigenvalues}
\begin{align}
   \bm{v}_{\pm}     &= C_{\pm}\left[\begin{array}{c}  \! i\Big(\gamma_e - \kappaL +  i2\Omega_0 \pm 2\zeta \Big)/4g \\  1 \end{array} \!\right] \eqlab{eigenvectors}\\
   \lambda_{\pm}    &= \frac12 \Big( -\frac{\gamma_e + \kappaL}{2} - i\Omega_0 \pm \zeta \Big) \eqlab{eigenvalues},
\end{align} 
\end{subequations}
where $C_{\pm}$ are normalization constants and we defined the complex frequency, $\zeta\equiv \zetaR + i\zetaI$, as
\begin{align}\eqlab{Omega}
   \zeta \equiv \sqrt{-4g^2 + \Big(\frac{\gamma_e\minus \kappaL}{2} + i\Omega_0\Big)^2   }.
\end{align} 
The coupling between cavity mode $\hat{b}$ and the TLE changes the eigenstates of the Hamiltonian into hybridized states where the excitation is in a superposition between being a photon in the cavity and an electron in the TLE (commonly known as dressed states). 
Let us assume that there are no decay mechanisms, $\kappa\equal\gamma_e\equal 0$. The dressed states then have energies
\begin{align}\eqlab{dressed state energy}
   \lambda_{\pm} &= -i\frac12 \Big( \Omega_0  \mp  \zetaI \Big) = -i\frac12 \Big( \Omega_0  \mp  \sqrt{4g^2 + \Omega_0^2}\Big) \approx   -i\frac{\Omega_0}{2}\bigg[ 1 \mp \Big(1+ \frac12\frac{4g^2}{\Omega_0^2} \Big) \bigg] ~\Rightarrow \nn\\
   \lambda_{+} &= i\frac{g^2}{\Omega_0},~~\text{and} ~~ \lambda_{-} = -i\Big(\Omega_0 +\frac{g^2}{\Omega_0}\Big), ~~\text{for}~~ g \ll \Omega_0.
\end{align} 
In~\secref{controlled-phase gate}, we found that an additional phase of $(g^2/\Omega_0)t$ must be applied to the control field, $\Lambda(t)$, to absorb a one-photon wave packet in the limit $g \ll \Omega_0$. Since the photon must be coupled into the dressed state of the TLE-cavity system, ~\eqref{dressed state energy} provides an alternative way of arriving at this phase factor.

With the initial condition $\psi_{01g}(0)\equal 1$ and $\psi_{00e}(0)\equal 0$, ~\eqref{solution free evolution} becomes
\begin{subequations}\eqlab{solution free evolution example}
\begin{align}
   \psi_{01g}(t)    &= e^{-i\frac12 \Omega_0 t} \Big[ \cos\!\Big(\frac12 t \sqrt{4g^2+\Omega_0^2}\Big) + i\frac{\Omega_0}{\sqrt{4g^2+\Omega_0^2}} \sin\!\Big(\frac12 t \sqrt{4g^2+\Omega_0^2}\Big)   \Big] \\
   \psi_{00e}(t)    &= -i \frac{2g}{\sqrt{4g^2+\Omega_0^2} }e^{-i\frac12 \Omega_0 t} \sin\!\Big(\frac12 t \sqrt{4g^2+\Omega_0^2}\Big) \eqlab{solution free evolution example psi00e}.
\end{align}
\end{subequations}
\eqref{solution free evolution example psi00e} shows that the occupation probability of the emitter has a maximum of
\begin{align}\eqlab{P_00e max}
   \bigg|\psi_{00e}\bigg(t=\frac{\pi}{\sqrt{4g^2+\Omega_0^2}}\bigg) \bigg|^2  &= \frac{4g^2}{ 4g^2 +\Omega_0^2}. 
\end{align}
\eqref{solution free evolution example} also shows that the effective Rabi frequency of oscillation between the TLE and cavity is $ \sqrt{4g^2+\Omega_0^2}$, which means that the control field can be used to adjust this oscillation. 

For a two-photon input state, all the analysis above is similar except for the replacement $g\rightarrow \sqrt{2} g$. The difference is phase accumulation between a one- and two-photon input state is from~\eqref{dressed state energy} 
%
\begin{align}\eqlab{TLE phase difference}
   \lambda_{+}(\text{2 photons}) - \lambda_{+}(\text{1 photon})  &= \sqrt{8g^2 + \Omega_0^2} -  \sqrt{4g^2 + \Omega_0^2},
\end{align} 
which is maximized for $\Omega_0\equal 0$. It should therefore be expected that the optimal control function $\Omega(t)$ has a large value in the absorption and emission stage, while it has a small value close to zero during the interaction stage. This is confirmed by the result shown in~\figref{gate dynamics example}(a).

%% file: app_dephasing.tex
\section{Two-Level Emitter Dephasing \applab{dephasing}}
%
The master equation describing dephasing noise for a two-level system, that is otherwise evolving under a Hamiltonian $\hat{H}$, is 
\begin{align}
    \dot{\hat{\rho}} = -\frac{i}{\hbar}[\hat{H},\hat{\rho}] -\gammaDP [ \hat{\rho} - \hat{z} \hat{\rho} \hat{z}] 
\end{align}
in which $\hat{z}$ is the Pauli $\hat{\sigma}_z$ operator~\cite{Jacobs14}. 

A Poisson process is a random process in which instantaneous events occur at random times, with a constant probability of occurrence per unit time~\cite{Jacobs10c}. In each infinitesimal timestep $dt$ the probability that an event occurs is $P(dt) = \gammaDP dt$. If $\tau$ is the time between two consecutive events, the distribution of $\tau$ is 
\begin{align}\eqlab{Poisson pdf}
    \mathcal{P}_{\rm{dp}}(\tau) = \gammaDP e^{-\gammaDP \tau} . 
\end{align}
We can use the Poisson process to perform a Monte-Carlo simulation for the dephasing master equation. We sample a set of times $\tau_{j}$ to tell us when the events occur. Between events we evolve the system as 
\begin{align}
    \frac{d}{dt} |\psi\rangle = -\frac{i}{\hbar} \hat{H} |\psi\rangle . 
\end{align}
When an event occurs we transform the state as 
\begin{align}
    |\psi\rangle \rightarrow \hat{z} |\psi\rangle . 
\end{align}

We can see that the above stochastic simulation reproduces the master equation in the following way. At each time-step $dt$ the evolution is a mixture of 
\begin{align}
    \hat{\rho} \rightarrow \hat{\rho} - \frac{i}{\hbar}[\hat{H},\hat{\rho}] dt 
\end{align}
with probability $P_0 = 1 - \gammaDP dt$, and 
\begin{align}
    \hat{\rho} \rightarrow  \hat{z} \hat{\rho} \hat{z} 
\end{align}
with probability $P_1 = \gammaDP dt$. The full (average) evolution is thus (to first-order in $dt$)
\begin{align}
    \hat{\rho} & \rightarrow (\hat{\rho} - \frac{i}{\hbar}[\hat{H},\hat{\rho}] dt)P_0 +  \hat{z}\hat{\rho} \hat{z} P_1  \nonumber \\
    & = \hat{\rho} - \gammaDP \hat{\rho} dt + \gammaDP \hat{z}\hat{\rho} \hat{z} dt  - \frac{i}{\hbar}[\hat{H},\hat{\rho}] dt \nonumber \\
    & = \hat{\rho} - \frac{i}{\hbar}[\hat{H},\hat{\rho}] dt - \gammaDP [\hat{\rho} - \hat{z} \hat{\rho} \hat{z}] dt 
\end{align}
giving the master equation above. 

The reason the simulation is so simple in this case is because the probability of a jump (in this case a $pi$ phase kick) is independent of the state $\hat{\rho}$, and because the kick is a unitary operation.  

The state fidelity is then given by
\begin{align}\label{eq:state fidelity full}
    \mathcal{F}_s \equiv \bra{\mu_0} \hat{\rho}_s \ket{\mu_0} = \frac{1}{N_{\rm{traj}}} \sum_i^{N_{\rm{traj}}}  \left| \int_{t_0}^{\Tend} \!\!\!\! \int_{t_0}^{\Tend} \!\!\!\!  \xi_{\rm{out}}^{(i)}(t_m, t_n) \xi_{\mu}(t_n)^{\!*} \xi_{\mu}(t_m)^{\!*}dt_m dt_n \right|^2,
\end{align} 
where $\xi_{\rm{out}}^{(i)}(t_m, t_n)$ is the output calculated in the $i$th Monte-Carlo trajectory. 


%% file: app_FoM_comparison.tex
\section{Mode Volumes \applab{mode volumes}}
\subsection{Second-Order Nonlinearity}
To extract the relevant figures of merit from literature on second-harmonic generation, we consider two different ways of arriving at a definition for the mode volume corresponding to a second-harmonic generation (SHG) interaction. One is based on classical perturbation theory and the other on quantum mechanical perturbation theory. We connect the two derivations by considering equations of motion for quantum operators and corresponding classical field amplitudes. In quantum mechanics, these are derived starting from the Hamiltonian 
%
\begin{align}\eqlab{SHG Hamiltonian}
    \hat{H}_{\rm{SHG}} = \hbar \chiSHG\big( \hat{b}\hat{b}\hat{c}^\dagger + \hat{b}^\dagger\hat{b}^\dagger\hat{c} \big).
\end{align}
%
Equations of motion for operators, $\hat{O}$, are found from
\begin{align}\eqlab{Heisenberg equation}
    \frac{d\hat{O}}{dt} = \frac{-i}{\hbar} \big[ \hat{O}, \hat{H}_{\rm{SHG}}\big].
\end{align}
Using~\eqref{SHG Hamiltonian}, we find the commutation relations
\begin{subequations}\eqlab{operator commutations}
\begin{align}
    \big[ \hat{b}, \hat{H}_{\rm{SHG}} \big] &= 2 \hbar \chiSHG \hat{b}^\dagger\hat{c} ~~\Rightarrow~~ \frac{d\hat{b}}{dt}  \propto -2i\chiSHG \hat{b}^\dagger\hat{c} \\
    \big[ \hat{c}, \hat{H}_{\rm{SHG}} \big] &= \hbar \chiSHG \hat{b}\hat{b}  ~~\Rightarrow~~ \frac{d\hat{c}}{dt}  \propto -i\chiSHG \hat{b}^2.
\end{align}
\end{subequations}
The classical limit is found by removing the $\hat{~}$ from the operators (we omit a detailed justification for this here)
\begin{align}\eqlab{classical eoms}
    \frac{d b}{dt}  \propto -2i\chiSHG b^{\!*} c \quad \text{and} \quad  \frac{d c}{dt}  \propto -i\chiSHG b^{2} ,
\end{align}
where $|b|^2$ and $|c|^2$ represent the number of photons in modes $\hat{b}$ and $\hat{c}$. In Ref.~\cite{Lin2016}, the same equations are written as
\begin{align}\eqlab{classical eoms Rodriguez}
    \frac{d a_1}{dt}  \propto -i\omega_1 \beta_1 a_1^* a_2 \quad \text{and} \quad  \frac{d a_2}{dt}  \propto -i\omega_2\beta_2 a_1^{2} ,
\end{align}
where $|a_1|^2$ and $|a_2|^2$ represent the energy in modes $a_1$ and $a_2$ while $\beta_2\equal \beta_1^*/2$ with 
\begin{align}\eqlab{beta1}
    \beta_1 = \frac14 \frac{\int d\bm{r} \epsilon_0 \sum_{ijk} \chi_{ijk}^{(2)}(\bm{r}) \big( E_{1,i}^*E_{2,j} E_{1,k}^* + E_{1,i}^*E_{1,j}^* E_{2,k} \big) }{\int d\bm{r} \epsilon_0\epsilon_1 | E_1|^2 \sqrt{\int d\bm{r} \epsilon_0\epsilon_2 | E_2|^2}}.
\end{align}
If we consider a material like LiNbO$_3$ with a maximum tensor component in the diagonal e.g. $\chi_{xxx}^{(2)}$, we have 
\begin{align}\eqlab{beta1 33}
    \beta_{1,33} = \frac24  \frac{\tilde{\chi}_{33}^{(2)}}{\sqrt{\epsilon_0}} \frac{\int d\bm{r}  \bar{\epsilon}(\bm{r})  E_{1,x}^*E_{1,x}^* E_{2,x}  }{\int d\bm{r} \epsilon_1 | E_1|^2 \sqrt{\int d\bm{r} \epsilon_2 | E_2|^2}} =  \frac24  \frac{\tilde{\chi}_{33}^{(2)}}{\sqrt{\epsilon_0}} \frac{\bar{\beta}_{33}^*}{\sqrt{\lambda_1^3}} ~~\Rightarrow ~ \beta_{2,33} = \frac14  \frac{\tilde{\chi}_{33}^{(2)}}{\sqrt{\epsilon_0 \lambda_1^3}}\bar{\beta}_{33},
\end{align}
where we used contracted notation, $\tilde{\chi}_{33}^{(2)} \equiv \chi_{xxx}^{(2)}$ (when $x$ is the extra-ordinary crystal axis), that is often used when Kleinman's symmetry condition is valid~\cite{BoydBook}. The normalized overlap is defined as
\begin{align}\eqlab{betabar 33}
    \bar{\beta}_{33} = \frac{\int d\bm{r} \bar{\epsilon}(\bm{r}) E_{1,x}^2 E_{2,x}^* }{\int d\bm{r} \epsilon_1 | E_{1}|^2 \sqrt{\int d\bm{r} \epsilon_2 | E_{2}|^2}} \sqrt{\lambda_1^3},
\end{align}
where $\bar{\epsilon}(\bm{r})$ is a function that equals 1 inside the nonlinear material and 0 outside. 
Inserting~\eqref{beta1 33} into~\eqref{classical eoms Rodriguez} and comparing to~\eqref{classical eoms}, we find
\begin{align}\eqlab{chiNL betabar relation}
    \chiNL \frac{1}{\sqrt{2\hbar\omega_1}} = \omega_2\beta_{2,33} = 2\omega_1 \frac14  \frac{\tilde{\chi}_{33}^{(2)}}{\sqrt{\epsilon_0 \lambda_1^3}}\bar{\beta}_{33}  ~~\Rightarrow ~~  \chiNL = \sqrt{\frac{\hbar \omega_1^3}{2\epsilon_0 \lambda_1^3}}\tilde{\chi}_{33}^{(2)} \bar{\beta}_{33}.
\end{align}
Note that we divided by $\sqrt{2\hbar\omega_1}$ on the left hand side because the normalization in~\eqref{classical eoms} is related to the number of photons and to the total energy in~\eqref{classical eoms Rodriguez}. In Ref.~\cite{krastanov2021room}, equation (38) defines the nonlinear coupling rate
\begin{align}\eqlab{chiNL Krastanov Vshg}
     \chiNL = \sqrt{\frac{\hbar \omega_1 2\omega_1^2}{8 \epsilon_0}} \tilde{\chi}_{33}^{(2)} \frac{1}{n^3 \sqrt{V_{\rm{shg}}}}.
\end{align}
If we assume $n^3\equal n_1^3 \equal \epsilon_1 \sqrt{\epsilon_2}$, then equation (39) in Ref.~\cite{krastanov2021room} reads
\begin{align}\eqlab{shg mode volume}
     \frac{1}{\overline{V}_{\rm{shg}}} \equiv \frac{1}{V_{\rm{shg}}}\left( \frac{\lambda_1}{n_1} \right)^{\!\!3} \!= \left| \frac{\int d\bm{r} \bar{\epsilon}(\bm{r}) E_{1,x}^2 E_{2,x}^* \sqrt{\lambda_1^3}}{\int d\bm{r} \epsilon_1 | E_{1}|^2 \sqrt{\int d\bm{r} \epsilon_2 | E_2|^2}} \right|^2 = \big|\bar{\beta}_{33}\big|^2 .
\end{align}
Inserting~\eqref{shg mode volume} into~\eqref{chiNL Krastanov Vshg}, we have 
\begin{align}\eqlab{chiNL Krastanov beta33}
     \chiNL = \sqrt{\frac{\hbar \omega_1^3}{4 \epsilon_0\lambda_1^3}} \tilde{\chi}_{33}^{(2)} \bar{\beta}_{33},
\end{align}
which differs from~\eqref{chiNL betabar relation} by a factor of $\sqrt{2}$. The definition of the SHG mode volume in terms of $\bar{\beta}$ in Ref.~\cite{Lin2016} was derived from the purely classical perturbation theory in Ref.~\cite{Rodriguez2007} using the electric field, whereas $\overline{V}_{\rm{shg}}$ in Ref.~\cite{krastanov2021room} was derived from a quantum mechanical perturbation theory laid out in Refs.~\cite{Sipe2004, Bhat2006, Quesada2017} using the electric displacement field. Ref.~\cite{Quesada2017} explains why the electric displacement field should be used and the difference between~\eqsref{chiNL betabar relation}{chiNL Krastanov beta33} may originate in the different quantization procedures. \\ 

In materials where other tensor components of $\chill_{ijk}$ are dominant, the definition of $V_{\rm{shg}}$ will differ from~\eqref{betabar 33}. To make sure all the contributions to the summation in~\eqref{beta1} are counted, we write it out
\begin{align}\eqlab{beta 1 sum}
     \begin{array}{rccccc} ijk: & E_{1,i}^*E_{2,j}E_{1,k}^* + E_{1,i}^*E_{1,j}^*E_{2,k} & & & & \\
     & & & & &\\
     xxx: ~yxx: ~zxx: & E_{1,x}^*E_{2,x}E_{1,x}^* + E_{1,x}^*E_{1,x}^*E_{2,x} & \quad: & E_{1,y}^*E_{2,x}E_{1,x}^* + E_{1,y}^*E_{1,x}^*E_{2,x} & \quad : & E_{1,z}^*E_{2,x}E_{1,x}^* + E_{1,z}^*E_{1,x}^*E_{2,x}\\
     xxy: ~yxy: ~zxy: & E_{1,x}^*E_{2,x}E_{1,y}^* + E_{1,x}^*E_{1,x}^*E_{2,y} &  \quad :& E_{1,y}^*E_{2,x}E_{1,y}^* + E_{1,y}^*E_{1,x}^*E_{2,y} & \quad :& E_{1,z}^*E_{2,x}E_{1,y}^* + E_{1,z}^*E_{1,x}^*E_{2,y}\\
     xxz: ~yxz: ~zxz: & E_{1,x}^*E_{2,x}E_{1,z}^* + E_{1,x}^*E_{1,x}^*E_{2,z} &  \quad :& E_{1,y}^*E_{2,x}E_{1,z}^* + E_{1,y}^*E_{1,x}^*E_{2,z} & \quad :& E_{1,z}^*E_{2,x}E_{1,z}^* + E_{1,z}^*E_{1,x}^*E_{2,z}\\
     xyx: ~yyx: ~zyx: & E_{1,x}^*E_{2,y}E_{1,x}^* + E_{1,x}^*E_{1,y}^*E_{2,x} & \quad: & E_{1,y}^*E_{2,y}E_{1,x}^* + E_{1,y}^*E_{1,y}^*E_{2,x} & \quad : & E_{1,z}^*E_{2,y}E_{1,x}^* + E_{1,z}^*E_{1,y}^*E_{2,x}\\
     xyy: ~yyy: ~zyy: & E_{1,x}^*E_{2,y}E_{1,y}^* + E_{1,x}^*E_{1,y}^*E_{2,y} & \quad: & E_{1,y}^*E_{2,y}E_{1,y}^* + E_{1,y}^*E_{1,y}^*E_{2,y} & \quad : & E_{1,z}^*E_{2,y}E_{1,y}^* + E_{1,z}^*E_{1,y}^*E_{2,y}\\
     xyz: ~yyz: ~zyz: & E_{1,x}^*E_{2,y}E_{1,z}^* + E_{1,x}^*E_{1,y}^*E_{2,z} & \quad: & E_{1,y}^*E_{2,y}E_{1,z}^* + E_{1,y}^*E_{1,y}^*E_{2,z} & \quad : & E_{1,z}^*E_{2,y}E_{1,z}^* + E_{1,z}^*E_{1,y}^*E_{2,z}\\
     xzx: ~yzx: ~zzx: & E_{1,x}^*E_{2,z}E_{1,x}^* + E_{1,x}^*E_{1,z}^*E_{2,x} & \quad: & E_{1,y}^*E_{2,z}E_{1,x}^* + E_{1,y}^*E_{1,z}^*E_{2,x} & \quad : & E_{1,z}^*E_{2,z}E_{1,x}^* + E_{1,z}^*E_{1,z}^*E_{2,x}\\
     xzy: ~yzy: ~zzy: & E_{1,x}^*E_{2,z}E_{1,y}^* + E_{1,x}^*E_{1,z}^*E_{2,y} & \quad: & E_{1,y}^*E_{2,z}E_{1,y}^* + E_{1,y}^*E_{1,z}^*E_{2,y} & \quad : & E_{1,z}^*E_{2,z}E_{1,y}^* + E_{1,z}^*E_{1,z}^*E_{2,y}\\
     xzz: ~yzz: ~zzz: & E_{1,x}^*E_{2,z}E_{1,z}^* + E_{1,x}^*E_{1,z}^*E_{2,z} & \quad: & E_{1,y}^*E_{2,z}E_{1,z}^* + E_{1,y}^*E_{1,z}^*E_{2,z} & \quad : & E_{1,z}^*E_{2,z}E_{1,z}^* + E_{1,z}^*E_{1,z}^*E_{2,z}
     \end{array}
\end{align}
In Ref.~\cite{Minkov2019}, the authors consider the  $\tilde{\chi}_{31}^{(2)}$ component of GaN, which obeys  $\tilde{\chi}_{31}^{(2)}\equal \chill_{xxz}\equal \chill_{xzx}\equal \chill_{yyz}\equal \chill_{yzy}$~\cite{Sanford2005}. Using~\eqref{beta 1 sum}, we can therefore write the normalized overlap as
\begin{align}\eqlab{beta1 13}
    \bar{\beta}_{1,13} = \frac24  \frac{\tilde{\chi}_{13}^{(2)}}{\sqrt{\epsilon_0}} \frac{\int d\bm{r} \bar{\epsilon}(\bm{r}) \big(E_{1,x}^2 E_{2,z}^* + E_{1,y}^2 E_{2,z}^* \big) }{\int d\bm{r} \epsilon_1 | E_{1}|^2 \sqrt{\int d\bm{r} \epsilon_2 | E_{2}|^2}} \sqrt{\lambda_1^3}.
\end{align}
Since~\eqref{beta1 13} has the same numeric pre-factor as~\eqref{beta1 33} we see that the nonlinear coupling rate simply is  
\begin{align}\eqlab{chiNL Minkov}
     \chiNL = \frac12 \sqrt{\frac{\hbar \omega_1^3}{ \epsilon_0\lambda_1^3}} \tilde{\chi}_{13}^{(2)} \bar{\beta}_{13}, ~~\text{with} ~~ \bar{\beta}_{13} = \frac{\int d\bm{r} \bar{\epsilon}(\bm{r}) \big(E_{1,x}^2 E_{2,z}^* + E_{1,y}^2 E_{2,z}^* \big) }{\int d\bm{r} \epsilon_1 | E_{1}|^2 \sqrt{\int d\bm{r} \epsilon_2 | E_{2}|^2}} \sqrt{\lambda_1^3}.
\end{align}
Note also that we used $\tilde{\chi}_{13}^{(2)}\equal 5.3\,$pm/V for GaN~\cite{Sanford2005} as well as $n\equal 2.32$ and $\lambda_1\equal 1300\,$nm~\cite{Minkov2019} in~\tabref{comparison table}.

\subsection{Third-Order Nonlinearity}
The protocol for a controlled-phase gate based on third-order nonlinearity in Refs.~\cite{Heuck2020, Heuck2020a} used a self-phase modulation (SPM) interaction. The mode volume in that case is defined as~\cite{Notomi2010}
\begin{align}\eqlab{spm mode volume}
     \frac{1}{V_{\rm{spm}}}  = \frac{\int d\bm{r} \epsilon_r^2 \bar{\epsilon}(\bm{r}) |E_1|^4 }{ \big( \int d\bm{r} \epsilon_r(\bm{r}) | E_1|^2 \big)^2}.
\end{align}
In Ref.~\cite{Choi2017}, the SPM  Hamiltonian is $H_{\rm{spm}} = \hbar \eta (\hat{n}-1)\hat{n}$ and the nonlinear rate is given by 
\begin{align}\eqlab{Choi spm rate}
     \eta = - \frac{3\hbar\omega^2 }{4\epsilon_0\epsilon_r^2} \frac{\chi^{(3)}}{V_{\rm{spm}}}.
\end{align}
Comparing with the definition of the SPM Hamiltonian in Ref.~\cite{Heuck2020} (see equation 13a), we see that $\chi_3\equal 4\eta$ and the nonlinear coupling rate is therefore
\begin{align}\eqlab{Heuck spm rate}
     \chi_3 = - \frac{3\hbar\omega^2}{\epsilon_0\epsilon_r^2} \frac{\chi^{(3)}}{V_{\rm{spm}}} = - \frac{3\hbar\omega^2}{\epsilon_0\epsilon_r^2} \left(\frac{n}{\lambda} \right)^{\!3} \frac{\chi^{(3)}}{\overline{V}_{\rm{spm}}} = - \frac{3\hbar\omega^2}{\epsilon_0 n \lambda^3} \frac{\chi^{(3)}}{\overline{V}_{\rm{spm}}} .
\end{align}
The cavity design in Ref.~\cite{Choi2017} had a $Q$ of $2\!\times\!10^6$ and $Q\lambda^3 /V_{\rm{spm}}=5\!\times\!10^8$, which leads to a mode volume of  $\overline{V}_{\rm{spm}}\equal 2\!\times\!10^6\!\times\! 3.48^3 / 5\!\times\!10^8 = 0.17 $ using $n\equal 3.48$.